\documentclass[aps,pre,twocolumn,superscriptaddress]{revtex4-1}  % 

\usepackage{graphicx}  % needed for figures
\usepackage{amssymb}   % for math
\usepackage{amsmath}
\usepackage{color}
\usepackage{times}

\begin{document}

\title{Avoidance, Adjacency, and Association in Distributed Systems Design}
\author{Andrei A.~Klishin}
\altaffiliation[Current address: ]{Department of Bioengineering, School of 
	Engineering \& Applied Science, University of Pennsylvania, Philadelphia, 
	PA 19104, USA}
\affiliation{Department of Physics, University of Michigan, Ann Arbor, Michigan 48109, 
USA}
\affiliation{Center for the Study of Complex Systems, University of Michigan, Ann Arbor, 
Michigan 48109, USA}
\affiliation{John A. Paulson School of Engineering and Applied Sciences, Harvard 
University, Cambridge, MA 02138, USA}
\author{David J.~Singer}
\affiliation{Naval Architecture and Marine Engineering, University of Michigan, Ann 
Arbor, Michigan 48109, USA}
\author{Greg van Anders}
\affiliation{Department of Physics, University of Michigan, Ann Arbor, Michigan 48109, 
USA}
\affiliation{Center for the Study of Complex Systems, University of Michigan, Ann Arbor, 
Michigan 48109, USA}
\affiliation{Department of Physics, Engineering Physics, and Astronomy, Queen's
University, Kingston, Ontario, K7L 3N6, Canada}
\email{gva@queensu.ca}

\date{\today}

\begin{abstract}
  Patterns of avoidance, adjacency, and association in complex systems design
  emerge from the system's underlying logical architecture (functional
  relationships among components) and physical architecture (component physical
  properties and spatial location). Understanding the physical--logical 
  architecture
  interplay that gives rise to patterns of arrangement requires a quantitative
  approach that bridges both descriptions. Here, we show that statistical
  physics reveals patterns of avoidance, adjacency, and association across sets
  of complex, distributed system design solutions. Using an example arrangement
  problem and tensor network methods, we identify several phenomena in complex
  systems design, including placement symmetry breaking, propagating
  correlation, and emergent localization. Our approach generalizes
  straightforwardly to a broad range of complex systems design settings where it
  can provide a platform for investigating basic design phenomena.
\end{abstract}

\maketitle
\begin{figure}
	\includegraphics[width=0.45\textwidth]{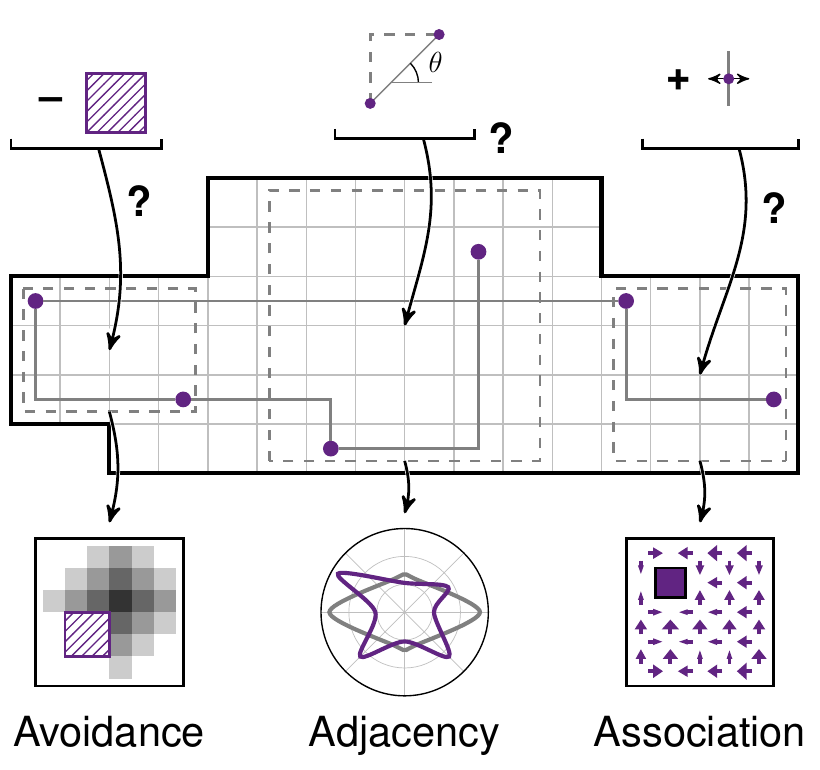}
	\caption{%
    Complex system design raises the question of identifying
    arrangement patterns of avoidance, adjacency, and association.
    Avoidance patterns (left) can be probed by testing the ``cost'' of
    creating a void in the design. Adjacency patterns (center) describe
    arrangement motifs found in the design, e.g.\ angles between the placement
    of design elements. Association patterns (right) relate to the preference
    for proximity between design elements, e.g.\ measures ``preferred''
    locations in adding design elements.
  }
	\label{fig:TripleA}
\end{figure}

A fundamental question in the design of complex, multicomponent systems is how
the components of the system are
arranged.\cite{aikens1985facility,dorneich1995global,drira2007facility}
Arrangement problems, generically, present the challenge of anticipating or
identifying ``prime real estate'',\cite{owen1998strategic, Shields2017} i.e.\
sectors of the system's architecture that have premium or priority because of
the mutual avoidance, adjacency, or association between system components (see
Fig.\ \ref{fig:TripleA}).  Determining or anticipating components' patterns of
avoidance, adjacency, and association is important in so-called ``greenfield''
settings, i.e.\ before design aspects have been specified, and in ``brownfield''
settings, i.e.\ when one or more system design aspects have been
determined.\cite{dorsey2003brownfields,adams2008greenfields,hopkins2008eating}
In both greenfield and brownfield settings, determining the patterns of
arrangement and identifying system factors driving those behaviors is crucial
for managing, mitigating, or adapting to likely design
outcomes.\cite{chalfant2015}

Managing likely design outcomes by identifying patterns of arrangement depends
crucially on both a system's logical architecture, i.e.\ the set of functional
connections between components, and on the system's physical architecture, i.e.\
the physical properties of the components and their arrangement in
space.\cite{Brefort2018} A system's logical architecture is essentially
topological, and can be treated using network theory
techniques.\cite{newman2010networks} In contrast, describing the physical
architecture of a system is typically done by disciplinary engineering
approaches that rest on known physical principles. Treating problems in
arrangements that arise from an interplay of a system's logical and physical
architecture requires a framework that bridges a system's network-theory-level
description and its physical/spatial description. Whereas approaches exist at
the network theory and at the physical spatial levels, how they can be bridged
is an open question.

Here, we show that topological and physical descriptions of complex systems
design can be bridged using statistical physics. Using statistical physics we
demonstrate a framework that reveals patterns of avoidance, adjacency, and
association in arrangement problems. We use an example arrangement problem to
concretely demonstrate how our framework can identify patterns of arrangement
and how those patterns are driven by the design's logical and physical
architecture, in both greenfield and brownfield settings.

\section{Systems Physics Framework}
\subsection{Motivation: Design Challenges Lurk Between Logical and Physical
Architectures}
Complex systems are typically comprised by several interacting entities. The
interactions among the entities are often described at two different levels: the
description of what-is-connected-to-what, which is mathematically a graph-theoretic
description, and by the description of how the entities physically interact with
one another in space, which is described by physics. Taken separately, both
levels of description give useful but incomplete insights into design.

The logical architecture's graph-theoretic description of a complex system
design is valuable because it isolates the connections between system elements
that underly functionality.\cite{Brefort2018} Functionality in the logical
architecture is reduced to the topology of connections, and this connection
topology can be analyzed with network theory
techniques.\cite{buldyrev2010catastrophic,Stanley2012failures,Shields2016}
Network theory approaches to analyzing logical architecture are powerful because
they abstract out the system's physical realization.\cite{newman2003structure}
However, realizing the logical architecture physically can produce emergent
functional connections that are lost when logical architecture is analyzed
alone.

The physical architecture describes the realization of a complex system design
in terms of physical entities with physical properties. Whereas entities in the
logical architecture have abstract interactions that are encoded topologically,
in the physical architecture interactions interact mechanically, thermodynamically,
electromagnetically, etc., depending on physical factors such as energy
consumption and proximity in space. By retaining this level of detail, the
physical architecture provides an intimate picture of the performance of design
elements. However, this intimate portrait of performance typically describes a
single physical architecture instance. What that single instance means for the
space of possible designs more generally is often unclear.

Though they can provide key insight into single design instances, both physical
and logical architecture descriptions restrict our ability to understand general
characteristics of design. This restriction exists because general design
characteristics are properties of design problem spaces rather than of design
instances.  The focus on design instances has been described previously as
design organized around ``product structures'', i.e., around a particular
outcome of the design process.\cite{shields2017emergent} Contrasting with
product structures are ``knowledge structures'' that organize the design process
around relationships between design elements that persist across
instances.\cite{shields2017thesis}

\begin{figure*}
	\includegraphics[width=\textwidth]{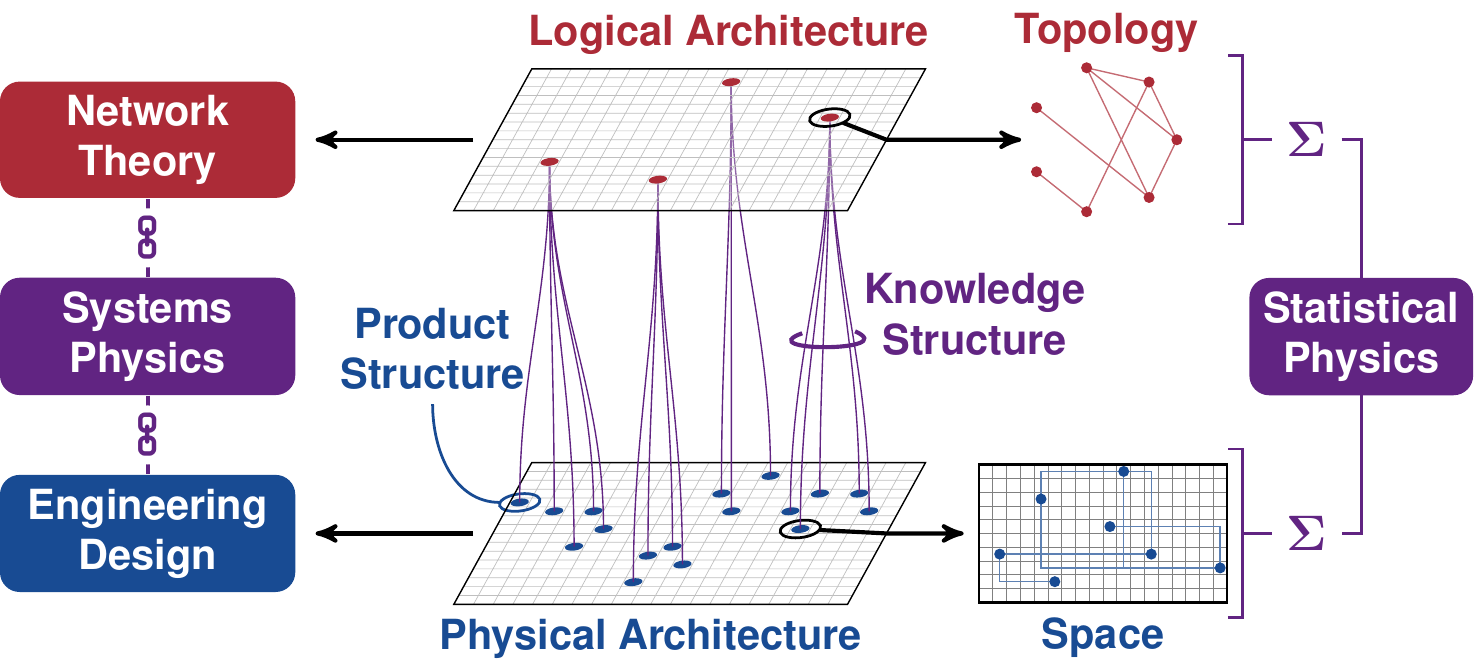}
	\caption{%
    The need to grow from ``product structure'' approaches, the focus on single
    design instances, to ``knowledge structure'' approaches, the patterns of
    design outcomes or challenges that persist across collections of instances,
    suggests the applying the framework statistical physics. Statistical
    physics collectively sums the topological-level description of the logical
    architecture that expresses the system's underlying functionality and the
    physical/spatial description of design instances. Connecting the logical and
    physical descriptions of design in this way, the resulting ``systems
    physics'' picture that emerges bridges between traditional network theory
    and engineering design approaches.
	}
	\label{fig:setup}
\end{figure*}

Searching across instances is key for identifying patterns of avoidance,
adjacency, and association that are generic features of design problem spaces.
Achieving this requires a different approach. To formulate this approach, the
key challenge is in framing knowledge that emerges from collections of possible
instances of a system. The problem of many instances that give rise to
collective behavior is the underlying principle that motivated the development
of statistical mechanics.\cite{gibbs} The fact that an analogous problem emerges
in design, i.e.\ the need to formulate knowledge structures to identify patterns
in design space, suggests that statistical physics could serve as the foundation
for a similar approach. Fig.\ \ref{fig:setup} illustrates this strategy of
attack.

\subsection{Statistical Physics Approach}
The need to address the problem of identifying patterns of avoidance, adjacency,
and association that persist across spaces of designs points to statistical
physics as a framework. To construct this framework there are two key
challenges: formulating the design problem as a statistical mechanics model,
and extracting from the model the knowledge structures that encode design space
properties.

To construct statistical physics models of design, we need two things:
the space of states and some metric on this space. For design problems that are
studied with optimization techniques like simulated annealing, these two things
are already at hand. A generic approach for constructing a statistical physics
design framework given a space of possible designs and a set of design
objectives was developed in Ref.\ \cite{systemphys}.

\emph{Mathematical Formulation}---Following Ref.\ \cite{systemphys}, we
denote the \emph{design space} as the set $\{\alpha\}$ consisting of individual
design solutions $\alpha$. Each
solution $\alpha$ can be quantitatively evaluated with a \emph{design objective}
$\mathcal{O}(\alpha)$. Instead of exclusively focusing on the design that
minimizes $\mathcal{O}$, we consider a probability distribution $p_\alpha$
over designs.\cite{systemphys} 
Of all possible normalized probability distributions, we seek one that 
	maximizes the Shannon entropy functional:\cite{jaynes1}
\begin{equation}
		S[p_\alpha]=-\sum\limits_{\alpha}p_\alpha \ln p_\alpha - \lambda 
		\left(\sum\limits_{\alpha} p_\alpha\mathcal{O}(\alpha) - 
		\left<\mathcal{O}\right> \right).
\end{equation}

We find the probability distribution by taking a functional derivative 
	$\delta S/\delta p_\alpha$ and setting it to zero, resulting in:
\begin{equation}
	p_\alpha = \frac{1}{\mathcal{Z}}e^{-\lambda\mathcal{O}(\alpha)};\quad 
	\mathcal{Z}=\sum\limits_{\alpha} e^{-\lambda\mathcal{O}(\alpha)},
	\label{eqn:boltzmann}
\end{equation}
where $\lambda$ is the \emph{design pressure}, or the relative importance of the
corresponding design objective in driving the distribution. The normalization
$\mathcal{Z}$ is known as the \emph{partition function} and contains a wealth of
information on the properties of the \emph{whole} design space. Mathematically, 
our study of the design problem is reduced to studying how the distribution 
$p_\alpha$ is affected by soft constraints (design pressure $\lambda$) and hard 
constraints (the set of available solutions $\alpha$).

\emph{Extracting Design Information}---With this formulation of design problems 
as statistical mechanics models, the
next challenge is to extract collective properties that encode information
about the structure of the design problem and the space it lives in. Doing this
typically requires computing sums over a combinatorially large set of $\alpha$,
which relies on various problem-specific mathematical techniques.
For simple scenarios involving only one or two nodes with integrable
interactions, prior work has shown that this can be done by coarse-graining to
extract effective, so-called Landau, free energies.\cite{robustdesign} We expect
that effective free energy approaches will, as they do in the ordinary
statistical mechanics of particles, provide the means to gain insight into the
collective properties of more complex design spaces. However for more complex
design spaces, again as is the case in the ordinary statistical mechanics of
particles, some mathematical techniques are required to study systems that lack
closed-form, integrable interactions.

To meet the challenge of extracting information about design spaces with complex
forms of interaction it is useful to take cues from the structure of the
problem. For complex problems the advantage of the logical architecture is that
it reduces the complexity of interactions among elements to simple, binary,
yes/no connections. The disadvantage of this simplicity is that it loses the
richness and specificity of the underlying design problem. This suggests a more
complete treatment of the design problem would be to ``decorate'' the
topological description of the logical architecture with information about the
``topography'' of the underlying design space and its physical architecture.
This topographic decoration can be carried out by encoding the design space as a
tensor network.

\emph{Tensor Network Formulation}---Tensor networks were originally introduced 
as a graphic notation for geometric
tensors,\cite{penrose1971} but over the last 25 years have grown into powerful
computational tools for storing and manipulating high-rank data. The tensor
network computations are especially efficient when the connections are sparse.
This property spurred the popularity of tensor networks in a broad range of
applications, from encoding entangled wavefunctions in quantum condensed matter
systems,\cite{verstraete2008matrix,schollwock2011density,orus2014practical} to
performing precision quantum chemistry calculations,\cite{chan2011density}
renormalizing lattice models,\cite{xie2009second,levin2007tensor} solving
constraint counting problems,\cite{biamonte2015tensor,kourtis2018fast}
accelerating numerical linear
algebra,\cite{cichocki2016tensor,cichocki2017tensor,oseledets2011tensor} and
learning multilinear classifiers in machine
learning.\cite{stoudenmire2016supervised} Across these applications, tensor
networks serve as an \emph{information structure} that contains an exhaustive
but raw description of the system.

We use tensor networks to bridge the logical and physical descriptions of a
design problem space. Network nodes encode the design elements' topography in
the physical space of their placement and their properties.
Network connections encode the functional connection topology and topography of
the physical interaction of design elements based on their spatial location and
physical properties. The topography--topology connection that the tensor network
encodes has the useful side benefits that it provides a simple graphical
representation of the interaction of design elements, and that well-developed
methods exist for extracting information from tensor networks.

Tensor networks encode information about the design space, but knowledge about
patterns of avoidance, adjacency, and association among design elements, has to
be extracted with special techniques. Extracting this information requires 
adding specifically
formulated pieces to it that represent key design questions (or, in physics
language, observables or order parameters). We formulate design questions
about avoidance, adjacency, and association among design elements by acting on
the tensor network with a combination of elementary ``moves''.  The moves yield
patterns of the placement over sets of solutions in design space, that are
computed via contraction of tensor networks. See Appendix~\ref{sec:methods} for 
a detailed
description of moves and contraction.

We note that recasting the problem in the language of tensor networks is 
an exact operation. The tensor networks shown on figures below are not 
qualitative illustrations, but mathematical formulas written in a graphical 
language explained in detail in Appendix~\ref{sec:methods}. Though there are 
many ways to perform and illustrate statistical mechanics computations for 
systems of simple topology (e.g.\ well-mixed, lattice, tree), to our knowledge 
tensor networks are the first method that can deal with arbitrary complex 
topology. Alternatively, we could evaluate the arrangement patterns by 
averaging over a representative sample of solutions with Monte Carlo methods, 
however that would engender statistical and sampling issues that could restrict
interpretability. Tensor network computations are free of those issues, at
least for the present problem. The numerical contraction 
computations introduce a controlled approximation error via SVD truncation, but 
it is of qualitatively different kind than finite sample error.

\subsection{Example Model System}
\begin{figure*}
	\begin{center}
		\includegraphics[width=.8\textwidth]{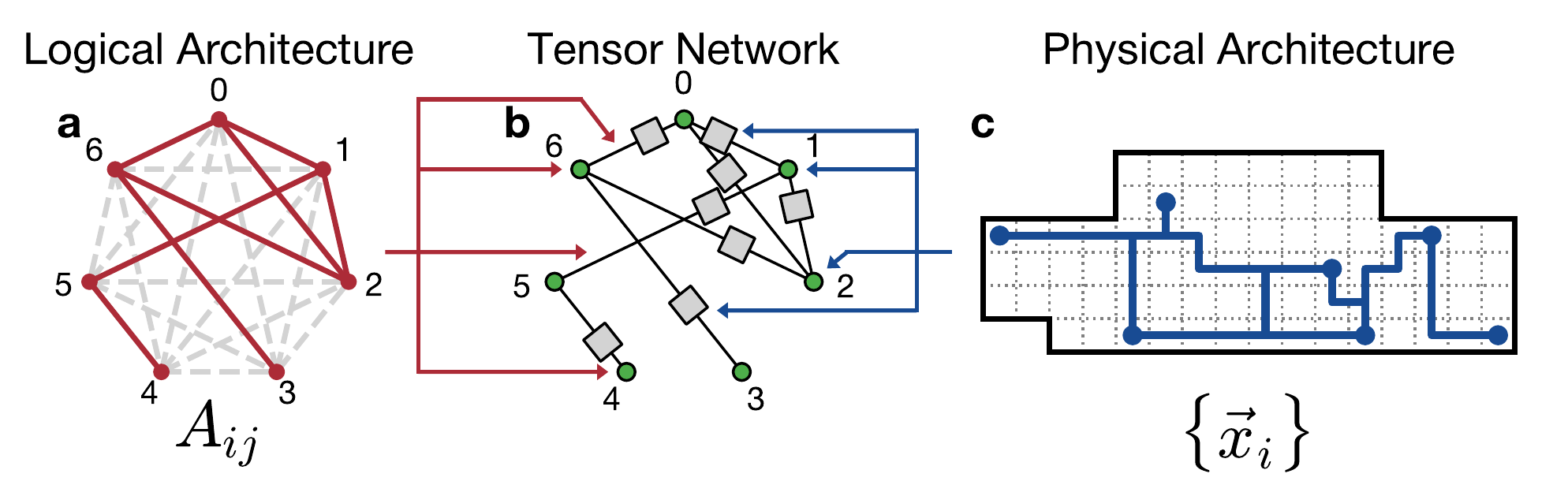}
	\end{center}
	\caption{Tensor network bridges the logical and physical descriptions of the design space for an example Naval Engineering arrangement problem. (a) The logical architecture is represented graphically by a network of seven functional units (red circles) and their specific pattern of functional connections (red lines), and algebraically with the adjacency matrix $A_{ij}$. The dashed gray lines represent the non-links in the network, which do not directly drive the arrangement but can be investigated. (b) The structure of the whole design space is contained in an information structure in form of a tensor network. Logical architecture determines the network pattern in which the tensors are connected, while physical architecture determines the contents of both site and coupling tensors. (c) The physical architecture is represented graphically by a square grid within a complex hull shape, and algebraically by the set of possible unit locations $\{\vec{x}_i\}$. A particular arrangement consists of the placement of all seven functional units within the hull and the routing of all functional connections between them (blue circles and lines). 
	}
	\label{fig:example}
\end{figure*}

\emph{Functional Units and Connections}---To demonstrate the Systems Physics
analysis and tensor network computations on a concrete example of a design
problem, we first define the specific logical and physical architectures for the
problem. We use a problem from Naval Engineering,\cite{Shields2017} in which the
functional units of a shipboard system need to be arranged within the hull of a
naval vessel, while respecting their functional connections, such as pipes
or cables. The network pattern of connections constitutes the logical
architecture, also represented algebraically as an adjacency matrix $A_{ij}$. We
find a wide variety of network motifs arise in networks of as few as $n=7$
functional units without any graph symmetries (Fig.~\ref{fig:example}a),
which we will study in the remainder of this work.

\emph{Ship Hull}---We position units and connections within a ship hull that we
represent, following \cite{Shields2017}, by a 2D square grid with a complex, but
fixed boundary (Fig.~\ref{fig:example}c). We constrain all connections
within the ship hull to always run along a shortest path between the two
functional units; we choose our hull to be $L_1$-convex to ensure that at least
one shortest path exists between any pair of cells. We find that hull models with a
few tens of cells are sufficient to establish placement patterns; computations
reported below are for hulls with $Y_0=78$ distinct cells for unit placement
each labelled as $\vec{x}_i$.

\emph{Design Objectives}---Picking the locations of all units and the routings
of functional connections between them defines a design solution
$\alpha=(\{\vec{x}_i\},\textsf{routing})$. In early-stage design, design
architectures are typically not fixed, therefore the full combinatorial design
space needs to be considered. Each design solution is quantitatively evaluated
with a \emph{design objective} $\mathcal{O}(\alpha)$, here we model routing
cost:
\begin{equation}
\lambda\mathcal{O}(\alpha)=\lambda\sum\limits_{ij}A_{ij}C L_1(\vec{x}_i,\vec{x}_j) \;,
\label{eqn:obj_fn}
\end{equation}
where $L_1$ is the ``Manhattan'' distance between the two cells, $C$ is the cost
per unit distance, and $\lambda$ is the design pressure. Given the placement of
units, we consider all allowed shortest paths between them. By definition, all
shortest paths have the same length, so the value of $\mathcal{O}$ doesn't
depend on the particular routing chosen, yet the number of routings is
important. To account for the redundancy of routings, we introduce an
\emph{effective design objective}:
\begin{align}
\lambda\mathcal{O}_\textsf{eff}(\{\vec{x}_i\})=&\sum\limits_{ij}A_{ij}
f(\vec{x}_i,\vec{x}_j;T)\; ; \\
f(\vec{x}_i,\vec{x}_j;T)=&\frac{C}{T}L_1(\vec{x}_i,\vec{x}_j)-\ln 
n_\textsf{rout}(\vec{x}_i,\vec{x}_j)\; ,
\label{eqn:obj_fn_spec}
\end{align}
where $n_\textsf{rout}$ is the number of shortest routings between $\vec{x}_i$
and $\vec{x}_j$ within the ship hull, typically growing with distance. The
routing lengths $L_1(\vec{x}_i,\vec{x}_j)$ and the number routings
$n_\textsf{rout}(\vec{x}_i,\vec{x}_j)$ are fully determined by the shape of the
hull and can be precomputed, stored as matrices, and scaled by the design
pressure as needed.

\emph{Tensor Network}---The representation of the design space in form of a 
tensor network depends on both logical and physical architectures 
(Fig.~\ref{fig:example}b). Logical architecture in form of the network $A_{ij}$ 
determines the pattern in which the site and coupling tensors are connected. 
Physical architecture determines the set of available locations for all units 
$\{\vec{x}_i\}$ that is used as index for all tensors. The effective design 
objective $f(\vec{x}_i,\vec{x}_j;T)$ determines the entries of the coupling 
tensor. See Appendix~\ref{sec:methods} for a detailed mathematical discussion.

\emph{Greenfield/Brownfield Settings}---In the above formulation, the design
space $\{\alpha\}$ of the problem is the space of all possible arrangements of
each functional unit $\{\vec{x}_i\}$, and is it necessary to establish a means
of distinguishing units with fixed and variable position. This distinction is
necessary because the formulation needs to address arrangement before or after
some of the units have been placed. We refer to situations in which no units
have fixed placement as greenfield settings. Greenfield settings are generically
associated with green color-coding in results figures that follow. Also, we
refer to situations in which one or more units have fixed locations as
brownfield settings. Brownfield settings are generically associated with brown
color-coding in results figures that follow. In Results figures that describe
brownfield settings that combine placed units with yet-to-be-placed units we
make a visual distinction between the two by color-coding placed units and their
effects brown and yet-to-be-placed units green.

\emph{Low Cost, High Flexibility, and Crossover Regimes}---The formulation of
design problems in terms of spaces of solutions weighted by objectives of the
form of Eq.\ \eqref{eqn:obj_fn_spec} has been studied in Ref.~\cite{systemphys}.
As in Ref.~\cite{systemphys} we expect that the choice of the design pressure
associated with each objective (general case: Eq.~\eqref{eqn:boltzmann}; this
model: Eq.~\eqref{eqn:obj_fn}) will have qualitatively distinct effects on design
outcomes.\cite{systemphys} To maximize generality, we study design pressures
that correspond to multiple behavioral regimes. We do this by first expressing
the design pressure via its inverse $\lambda=1/T$, where $T$ is the \emph{cost
tolerance}. Low cost tolerance means that minimizing the routing cost
$\mathcal{O}$ is a strong driver of a design solution choice, whereas high cost
tolerance means that the choice among the design solution is not driven by
cost. Ref.~\cite{systemphys} showed that the system driven by this design
objective undergoes a large-scale rearrangement (akin to a phase transition, but
at finite-size) around $T_\textsf{crit}=C/\ln 2\approx 1.44 C$. We pick $C=1$ to
fix the measurement units for $T$. $T<T_\textsf{crit}$ favors low cost and we
therefore expect units to organize into motifs that facilitate short (cheap)
connecting paths. We expect this setting to be characterized by effective
attraction. $T>T_\textsf{crit}$ favors maximal flexibility and we
expect units to organize into motifs that facilitate maximizing routing
degeneracy. We expect this setting to be characterized by effective repulsion.
We expect that for $T\approx T_\textsf{crit}$ where cost and flexibility
drivers are competing on near-equal footing there will be a crossover in
behavior.

\section{Results}
	Having formulated our statistical mechanics approach and introduced the 
	model system, we turn to the arrangement patterns of distributed design 
	elements
	depicted in Fig.~\ref{fig:TripleA} and the quantitative drivers of
	the patterns of avoidance, adjacency, and association. We find that all 
	three
	patterns manifest behaviours that are analogous to behaviours observed in
	condensed matter physics.

	The avoidance patterns we operationalize here describe the
	propensity for functional units in design to preferentially avoid certain
	regions of space. In our example system, we test this by introducing a void
	within the ship hull and determining the effect of the void placement on the
	design objectives.

	The adjacency patterns we operationalize here describe the
	propensity for design elements to arrange into characteristic relative
	arrangements, or motifs, that persist despite the absolute placement of the
	elements. In our example system, we compute the distribution 
	of pairwise relative directions between the units and study the effects of 
	topological distance and topology change.

	The association patterns we operationalize here describe the propensity for 
	an
	element introduced into a design to locate preferentially relative to the
	placement of already-existing functional units. In our example system, we
	compute the how placement affects design objectives, and find effective 
	forces
	that drive individual units to preferred locations in a later stage of
	design.

	Our results for each of these three arrangement patterns are described 
	below.

\subsection{Avoidance}
\begin{figure*}
	\includegraphics[width=.8\textwidth]{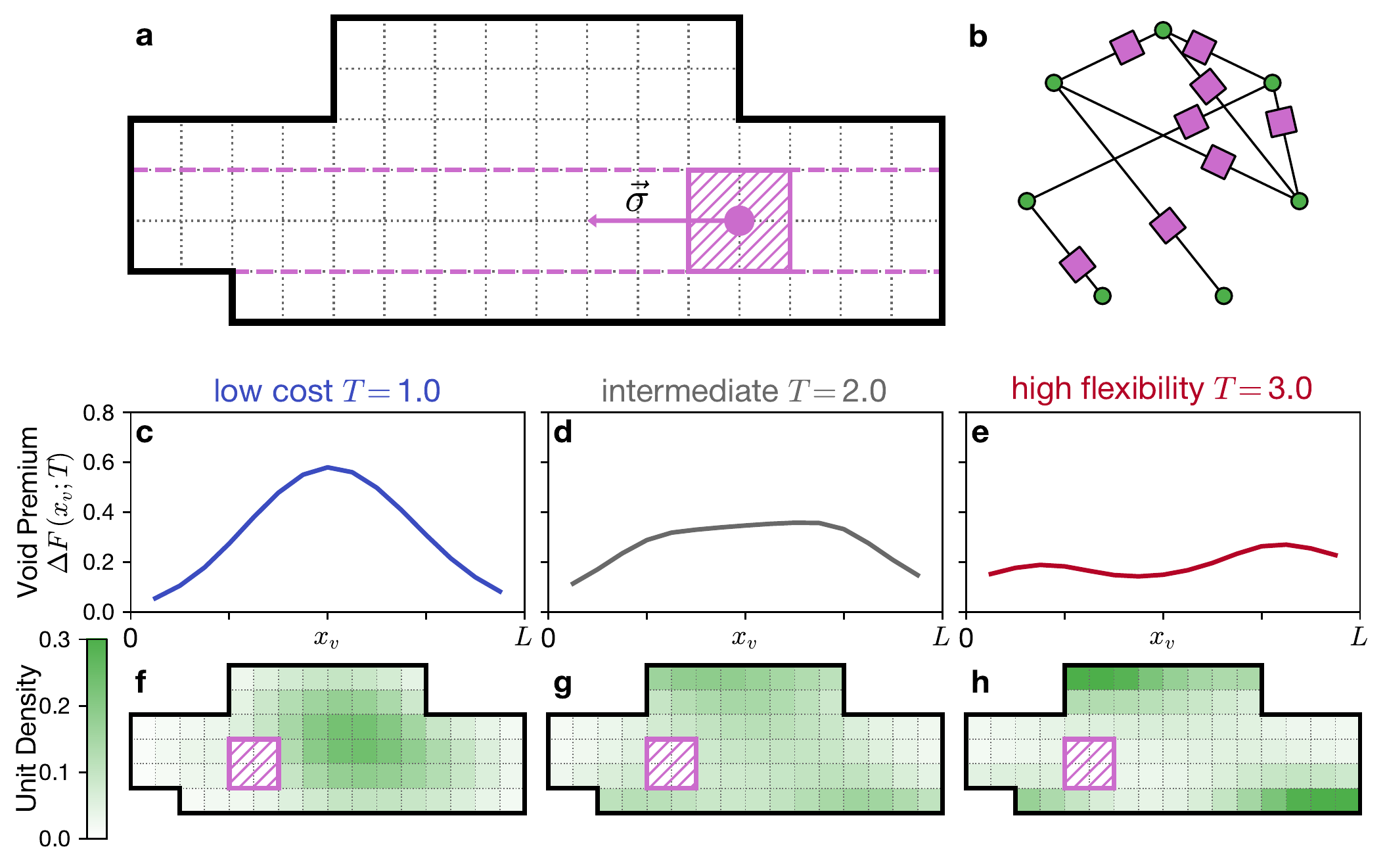}
  \caption{Void premium quantifies the cost of avoidance of reserved space
    across the whole design space in a greenfield scenario. (a) Schematic of the ship hull and square
    cells within (Physical Architecture). Pink square represents a void where
    unit placement is prohibited, driven by the void design stress
    $\vec{\sigma}$ along the center line of the hull (pink dashed lines). (b)
  Tensor network used to compute the void premium, with each coupling tensor
modified. (c-e) Graphs of void premium (void free energy $\Delta F(x_v;T)$)
against the void coordinate $x_v$ for three values of $T$ (color coded). (f-g)
Functional unit density in presence of the void.}
	\label{fig:void1}
\end{figure*}

\begin{figure*}
	\includegraphics[width=.8\textwidth]{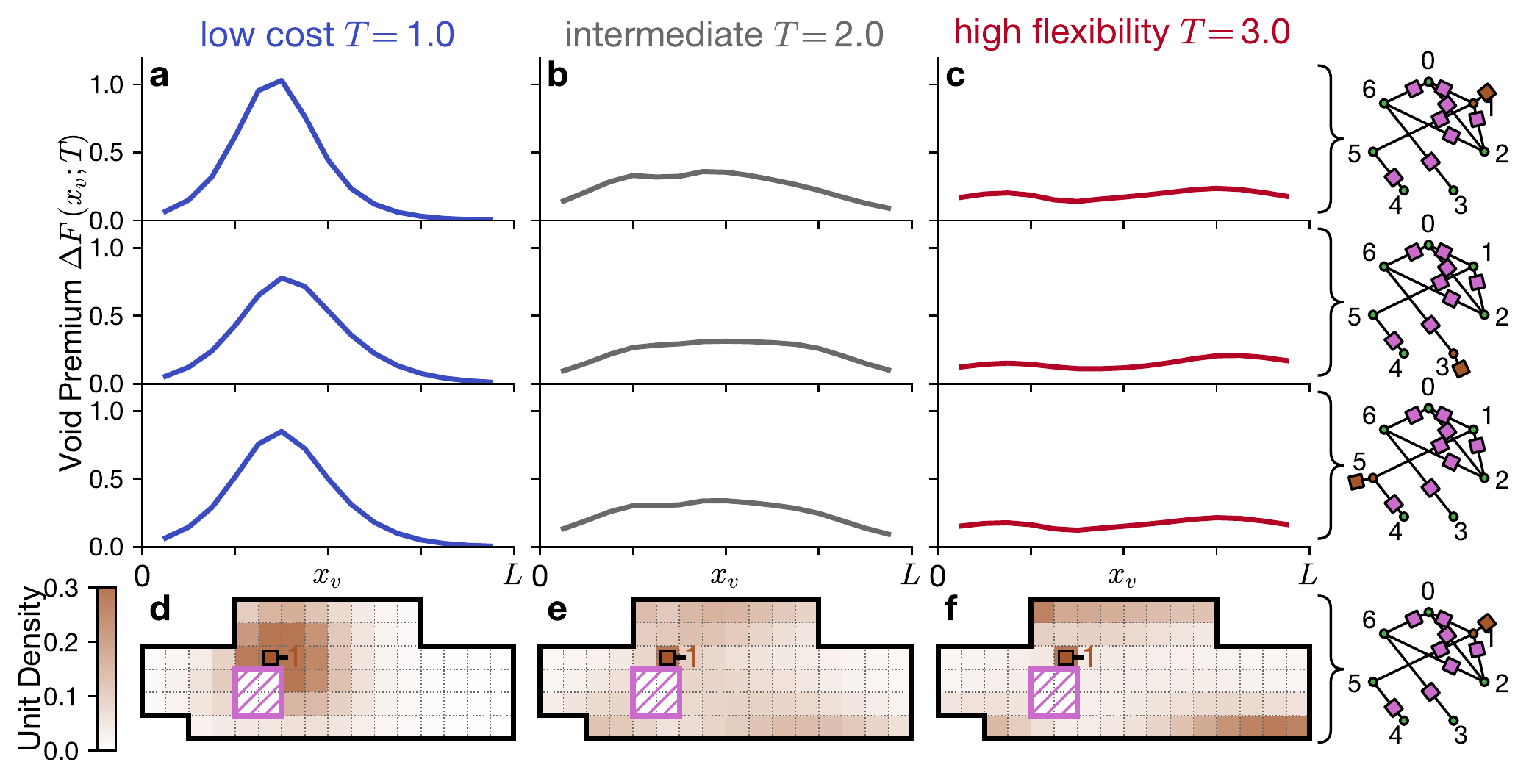}
	\caption{A brownfield scenario, such as anchoring one functional unit, sharpens the void premium curve. (a-c) Graphs of void premium (void free energy $\Delta F(x_v;T)$) against the void coordinate $x_v$ for three values of $T$ (columns, color coded) and different choice of the anchored unit (rows, anchor shown in the tensor network on the right). (d-e) Functional unit density in presence of the void and unit 1 anchored in the indicated cell (brown square).}
	\label{fig:void2}
\end{figure*}

\emph{Void Premium}---The interplay of logical and physical constraints among
design elements induces a complex landscape for element placement. Intervening
in that landscape by reserving space for future use could induce functional
units to make a complex, collective rearrangement to avoid the reserved space.
We characterize the cost of avoidance by computing the \emph{void premium} that
must be paid to forbid any units to be placed in the reserved space.

We implement reserved space mathematically by creating a \emph{void} $2\times 2$
cells in size. In the results below we introduce the void on the hull midline,
though in general it could be placed anywhere. We vary the horizontal location
of the void, $x_v$, from zero to the hull length $L$ (see
Fig.~\ref{fig:void1}a). We compute the cost of the void via the tensor network
approach by suppressing several rows and columns of the coupling
tensor (see Appendix~\ref{sec:methods}). Contracting the modified tensor 
network results in a
modified partition function, which is smaller or equal to the original one
$\mathcal{Z}(x_v;T)\le \mathcal{Z}(T)$. The ratio of the two partition functions
defines the non-negative free energy:
\begin{equation}
\Delta F(x_v;T)=-\ln \frac{\mathcal{Z}(x_v;T)}{\mathcal{Z}(T)}.
\label{eqn:void_cost}
\end{equation}
We take this void free energy as a measure of the void premium, the effective
``cost'' of the avoidance of a specified region in space.

\emph{Void Design Stress}---The magnitude of the void premium $\Delta F$
corresponds to the placement opportunity cost for the functional units. To
understand this opportunity cost, note that functional units that are not yet
placed form a greenfield ``cloud'' of possibilities within the hull, the
location and density of which depends on the cost tolerance $T$. Cutting a void
from a dense part of the cloud costs a lot of free energy, whereas cutting a
cloud from a sparse part of the cloud costs almost nothing. In this way,
scanning the void free energy along the void coordinate $x_v$ gives a direct
probe of the morphology of the cloud. Conversely, if we regard the unit
positions as fixed, and the void as moveable, the cloud of units drives the void
with an effective force $\sigma=-\Delta F/\Delta x_v$, which we call \emph{void
design stress}. The void free energy and design stress are then a concise
description of the collective effect of avoidance in functional unit placement.

The void premium and void design stress give a description of collective
avoidance effects in placement. These effects can be examined in greenfield and
brownfield settings.

\emph{Greenfield}---We studied avoidance metrics in greenfield settings, i.e.\
before any unit data have been fixed, in three design regimes specified by
cost tolerance $T$. We plot results in Fig.~\ref{fig:void1}c-e. At subcritical
$T=1.0$ (low cost priority), the void free energy curve shows a clear single
maximum in the middle of the hull, and two minima on the ends of the hull. At
near-critical $T=2.0$ (cost-flexibility tradeoff) the curve maintains the same
qualitative shape, but the maximum gets flatter. At supercritical $T=3.0$ (high
flexibility priority), the curve shape flips to have a local minimum at the
center of the hull, surrounded by local maxima on two sides. In other words, at
low $T$ the void prefers to be at either of the two ends of the ship (but a
choice needs to be made in favor of one of them). In contrast, at high $T$ the
void prefers to be in the center of the ship. Thus the change from designing for
flexibility (high $T$) to designing for cost (low $T$) induces a change from one
architecture class (central-void) to two architecture classes (bow-void and
stern-void). This collective effect is analogous to symmetry-breaking phase
transitions in conventional physical systems.\cite{goldenfeld}

To understand the origin of the symmetry breaking we note that void free
energy is a proxy for the morphology of the unit cloud. To illustrate the
shape of the cloud in a different way, we approximate the cloud density
as a sum of one-unit densities $\rho(\vec{x})=\sum_{i}p_i(\vec{x})$
and plot densities as heatmaps in Fig.~\ref{fig:void2}f-h. These heatmaps are
approximate because, unlike the void free energy curves, they ignore the
correlations in unit placement. At low $T$ (panel f), units attract each other
and thus preferentially form a cloud in the center of the hull and push
the void to either side of the hull. At near-critical $T$ (panel g), the
distribution becomes more homogeneous throughout the hull, flattening the curve.
At high $T$ (panel h), functional units strongly repel one another,
concentrating near the edges of the hull. This leaves the center nearly empty,
resulting in a single void free energy minimum.

\emph{Brownfield}---Both the void free energy curve and the unit cloud
morphology can, however, change dramatically in brownfield settings, e.g.\ if
even one unit is fixed to a specific location in space. We pick the location
indicated with the brown square in Fig.~\ref{fig:void2}d-f and fix one unit
there. We choose three different units to fix: unit 3 (which has 1
functional connection), unit 5 (2 functional connections) and unit 1 (3
functional connections). We plot the resulting void free energy curves in panels
a-c, and unit clouds in panels d-f.

Consider first the low-cost regime $T=1.0$ (Fig.~\ref{fig:void2}a,d). In the
greenfield setting (Fig.~\ref{fig:void1}f) units positions were determined
solely by ship geometry, and formed a dense cloud in the middle of
the ship (Fig.~\ref{fig:void1}f). Fixing a unit position places an additional
constraint on unit positions, and forces the unit cloud to condense around it
(Fig.~\ref{fig:void2}d). Because of this condensation, the void free
energy curve becomes simultaneously steeper and more focused around the fixed
unit point (panel a), but decays faster close to the edges of the hull. The void
free energy cost also depends on the topological position of the anchored unit:
it is highest for the most-connected unit 1 (bottom curve) and lowest for the
least-connected unit 3 (top curve).

At the near-critical and supercritical $T=2.0,3.0$ the anchored unit similarly
creates a reference point for the cloud, but the units in the cloud \emph{repel}
from that point. When repulsion and attraction are nearly balanced at $T=2.0$,
the cloud profile becomes nearly uniform and is not strongly affected by the
fixed unit (compare Fig.~\ref{fig:void1}g and Fig.~\ref{fig:void2}e). Similarly,
fixing a unit at supercritical $T=3.0$ creates a point of strong repulsion, forcing
the unit cloud to the opposite corners of the ship hull (Fig.~\ref{fig:void1}h
and Fig.~\ref{fig:void2}f). At both values of $T=2.0,3.0$, the cloud morphology
is not affected strongly by the single fixed unit position, and thus the
brownfield void premiums (Fig.~\ref{fig:void2}b-c) closely resemble their
greenfield couterparts (Fig.~\ref{fig:void1}b-c). Discussion of the effects of
avoidance on unit positions, i.e.\ backreactions on the cloud, can be found in
Appendix~\ref{sec:supres}.

\emph{Avoidance: Logical--Physical Architecture Interplay}---The above avoidance
analysis gives a case study of basic phenomenology of the interplay between
design pressure (favoring low-cost vs high-flexibility) and the logical and
physical architecture. Shifting the design priority from low-cost to
high-flexibility changed the interaction between pairs of functional units.
However, unit interactions were modulated by connection topology (i.e., logical
architecture) and by the spatial domain (i.e., the physical architecture). We
captured the effect of these complex interactions on spatial avoidance the unit
clouds in Figs.~\ref{fig:void1},\ref{fig:void2}. However, within the unit
clouds, the interplay of design pressure with the logical and physical
architectures also induces emergent coupling. This emergent coupling within the
cloud induces patterns of adjacency and association between units, which we turn
to next.

\subsection{Adjacency}
\begin{figure*}
	\includegraphics[width=.8\textwidth]{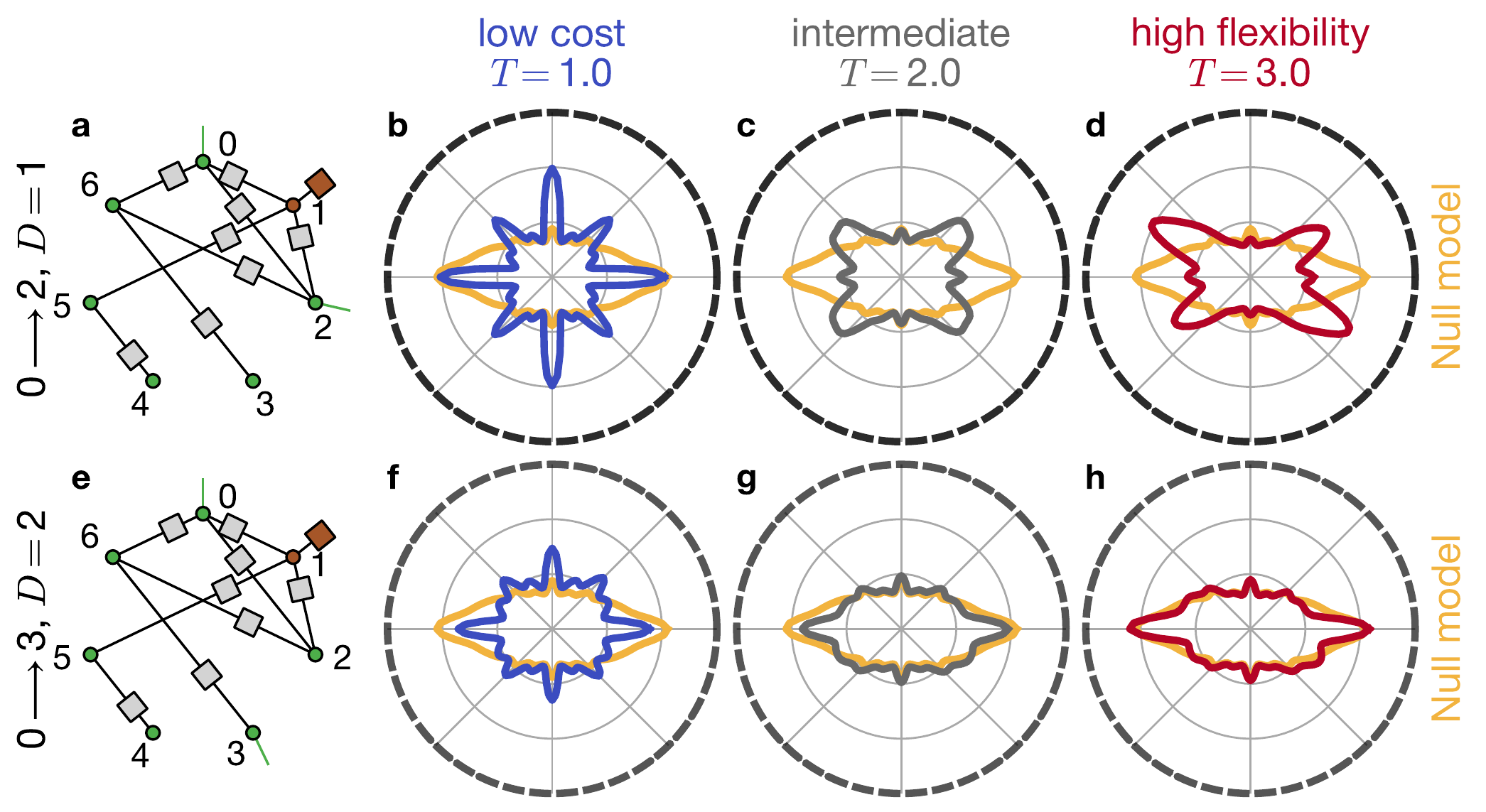}
	\caption{Bond diagrams quantify the adjacency patterns for both directly and indirectly connected functional units. (a-b) Tensor networks used for computations of bond diagrams for two pairs of units, one connected directly (0$\to$2, $D=1$), the other indirectly (0$\to$3, $D=2$). The site tensors in the measured pair have external legs (green). The unit 1 is anchored in the center of the hull (brown square). (c-h) Bond diagrams for the angle between the two units of the pair, for two different pairs (rows) and three values of $T$ (columns, color coded). Yellow curve shows the null model of the bond diagram, identical for all graphs. Black rim of bond diagram axes indicated the topological distance $D=1$, gray rim indicates $D=2$.}
	\label{fig:bonds1}
\end{figure*}

\begin{figure*}
	\includegraphics[width=.8\textwidth]{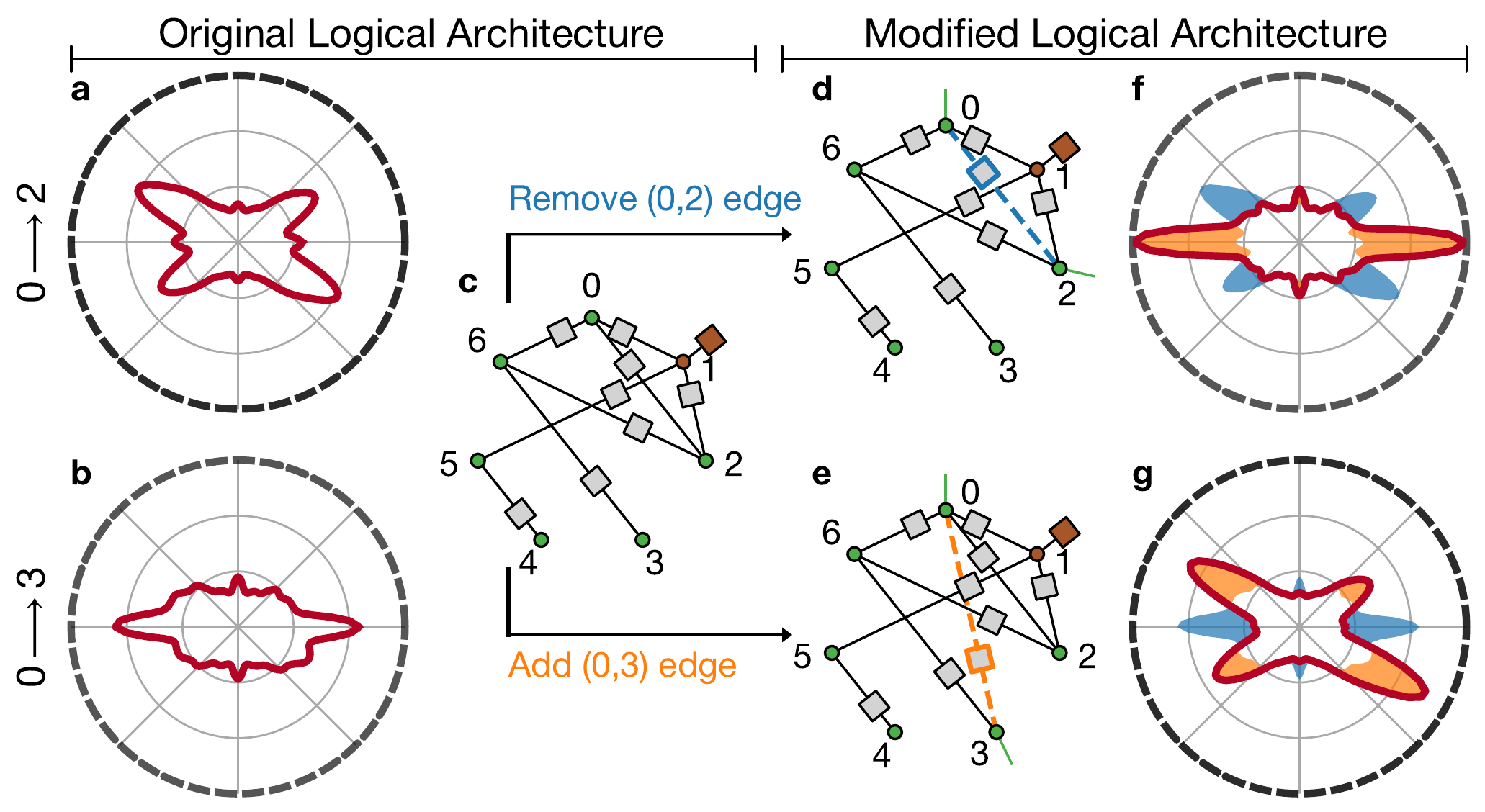}
	\caption{Small changes of Logical Architecture lead to large changes of adjacency patterns, visible in the difference of bond diagrams. Panels a-c correspond to the original Logical Architecture, while panels d-g illustrate two different changes of Logical Architecture. (a-b) Bond diagrams at $T=3.0$ for two pairs of units, 0$\to$2 ($D=1$) and 0$\to$3 ($D=2$). (c) Tensor network used for computations of bond diagrams with original Logical Architecture, external legs not shown. (d) Tensor network used for computation of the 0$\to$2 bond diagram when the functional connection (0,2) is removed (blue dashed line). (e) Tensor network used for computation of the 0$\to$3 bond diagram when the functional connection (0,3) is added (orange dashed line). (f-g) Resulting bond diagrams (dark red curve) with highlighted positive difference (orange shading) and negative difference (blue shading) with respect to the bond diagrams at the original Logical Architecture (panels a-b). Black rim of bond diagram axes indicated the topological distance $D=1$, gray rim indicates $D=2$.}
	\label{fig:bonds2}
\end{figure*}

Whereas our avoidance analysis derives from and illustrates the basic morphology
of the unit cloud within the ship hull, questions about unit adjacency derive
from correlations within the cloud, and the emergent coupling between units that
arise.

\emph{Bond-Diagram Measures of Adjacency}---To determine how emergent coupling
between units leads to arrangment motifs, we consider pairs of units and their
relative positions. We examine pairs of units in 2D space, and express motifs as
the polar angles $\theta (i\to j)$, which vary along with the positions of the
units in the cloud. Across the cloud, the angle takes form of a probability
distribution $p_{i\to	j}(\theta)$. This distribution is analogous to the bond
order that is used to describe structure in condensed
matter.\cite{steinhart,bondorder2000} In condensed matter, bond angles are
whole-system aggregate measures of adjacency. Here, the heterogeneous
connectivity of design elements yields ``bond'' diagrams that are specific to
each pair of units $i,j$.  Depending on whether the units $i$ and $j$ are
directly connected or not ($A_{ij}=1\;\text{or}\; 0$), the bond diagrams
illuminate the strength of direct or emergent adjacency patterns.

\emph{Computational Approach}---To compute the bond diagrams mathematically, we
use tensor networks to compute the raw 2-unit marginal distributions
$p(\vec{x}_i,\vec{x}_j)$ (Fig.~\ref{fig:bonds1}a,e), and convert them into the
angular distributions $p_{i\to j}(\theta)$ using Kernel Density Estimation to reduce the numerical artifacts (see
Appendix~\ref{sec:methods}).
In order to demonstrate more sharply defined bond diagrams, we assume that one
unit has already been placed (anchored) in the center of the hull and all other
units need to be placed with respect to it.
The bond diagram $p_{i\to j}(\theta)$ of any pair of units is not uniform with
respect to the angle $\theta$ even if the units are not connected at all,
directly or indirectly. This non-uniformity is driven by the shape of the hull,
a manifestation of the physical architecture, and we account for it by computing
\emph{null model} $p_0(\theta)$ of the bond diagram (see 
Appendix~\ref{sec:methods}). Differences
between the null model and computed bond diagrams are indicators of
interaction-driven adjacency. This interaction driven adjacency depends on unit
connectivity; we use a \emph{topological distance}, $D(i,j)$ metric. $D(i,j)$ is
the minimal number of network hops to get from unit $i$ to unit $j$. In our
example problem, the minimal number of hops varies from 1 (e.g.\ units 0$\to$1)
to 5 (e.g.\ units 3$\to$4).

\emph{Direct Adjacency}---We show topological distance, bond diagrams, and the
null model for our model in Fig.~\ref{fig:bonds1} in form of polar
plots for two example unit pairs (corresponding plots for all unit pairs are
given in Appendix~\ref{sec:supres}). We start discussion with the bond diagrams 
for direct adjacency
(0$\to$2, $D=1$). At subcritical $T=1.0$ (panel b) most units are located very
close to each other, either in cardinal or intercardinal directions
(orthogonally or diagonally), resulting in a bond diagram with a strong
eightfold signal. At near-critical $T=2.0$ (panel c) the orthogonal attraction
is balanced with diagonal repulsion, resulting in a bond diagram with smaller
peaks. At supercritical $T=3.0$ (panel d) the units are located relatively far
from each other and prefer diagonal relative location (since diagonal location
allows them to maximize their routing entropy), resulting in a fourfold,
X-shaped signal. The symmetry of the fourfold signal is further broken by
anchoring the unit 1.  This additional symmetry breaking is driven by the high
density of units in top-left and bottom-right corners of the hull (see
Fig.~\ref{fig:void1}).

\emph{Emergent Adjacency}---Across the whole $T$ range, the bond diagrams for
direct adjacency are significantly different from the null model. However,
adjacency can also be induced for indirectly connected unit pairs. The
indirectly connected unit pair 0$\to$3 shows emergent adjacency, since its bond
diagram is different from both the direct adjacency and the null model
(Fig.~\ref{fig:bonds1}f-h). The bond diagrams for unit pairs with even larger
topological distance the bond diagrams gradually approach the null model (see
Appendix~\ref{sec:supres}), following the intuition of decay of correlation 
functions with distance in
condensed matter systems. This observation suggests the general takeaway: at
large topological distance bond diagrams always approach the null model, and
thus are fully determined by the Physical Architecture; at small topological
distance the bond diagrams depend strongly on both the design pressure $T$ and
the explicit details of Logical Architecture. However, our results reveal that
for emergent adjacency topological distance is a good predictor of
\emph{strength} but is not a good predictor of \emph{shape}.

\emph{Logical Architecture Modifications}---To further test the interplay
between Logical Archiecture and Adjacency, we investigate what happens to bond
diagrams when we modify the Logical Architecture.  We consider two types of
modifications: removing an existing functional connection
(Fig.~\ref{fig:bonds2}d), or adding a new one between two units
(Fig.~\ref{fig:bonds2}e). Instead of comparing the
resulting bond diagrams to the null model again, we focus on the difference
between bond diagrams before and after modification (panels f-g).
\emph{Addition:} The effect of adding the (0,3) connection is strong. We can
anticipate that with the added direct adjacency, the adjacency pattern should
approach that of other directly adjacent units. The fourfold signal that results
(panel g) is a signal of this.  Since neither of the units 0 or 3 is explicitly
anchored in space, at high $T=3.0$ they want to be positioned at the opposite
ends of the longest diagonal available within the hull, in this case the
diagonal from top-left to bottom-right corners, similarly to the original
adjacency 0$\to$3 (Fig.~\ref{fig:bonds1}d).
\emph{Removal:} Like for addition the effect of removing the (0,2) connection is
dramatic, however the result is unexpected. Instead of fourfold diagonal direct
adjacency, the two units now have twofold horizontal emergent adjacency (panel
f). The reason for this is that with the (0,2) connection removed, the units
1-2-6-0 now form a rhombus.  Unit 1 is fixed in space, and because of high
$T=3.0$ all unit pairs prefer to have diagonal adjacency. In this case the units
0 and 2 on the opposite corners of the rhombus will have an orthogonal
adjacency, of which only horizontal adjacency manifests because the ship hull is
larger in length than height.

\emph{Adjacency: Changes and Constraints Drive Patterns}---We showed that the
irregular, complex-network nature of a system's Logical Architecture drives the
patterns of direct and emergent adjacency. We showed adjacency patterns can
change significantly with changes in Logical Architecture. One outcome of this
approach was the ability to detect emergent adjacency. The emergent effects we
observed were with a single fixed unit. Though having having few fixed units is
a characteristic of early-stage design, later-stage design situations will
result in more fixed units. Fixing more units will induce more constraints, and
further constraints will complicate the interplay of the logical and physical
architecture. A more complicated logical--physical architecture interplay should
induce more complex patterns of association between units, which we will examine
next.

\subsection{Association}
\begin{figure*}
	\includegraphics[width=\textwidth]{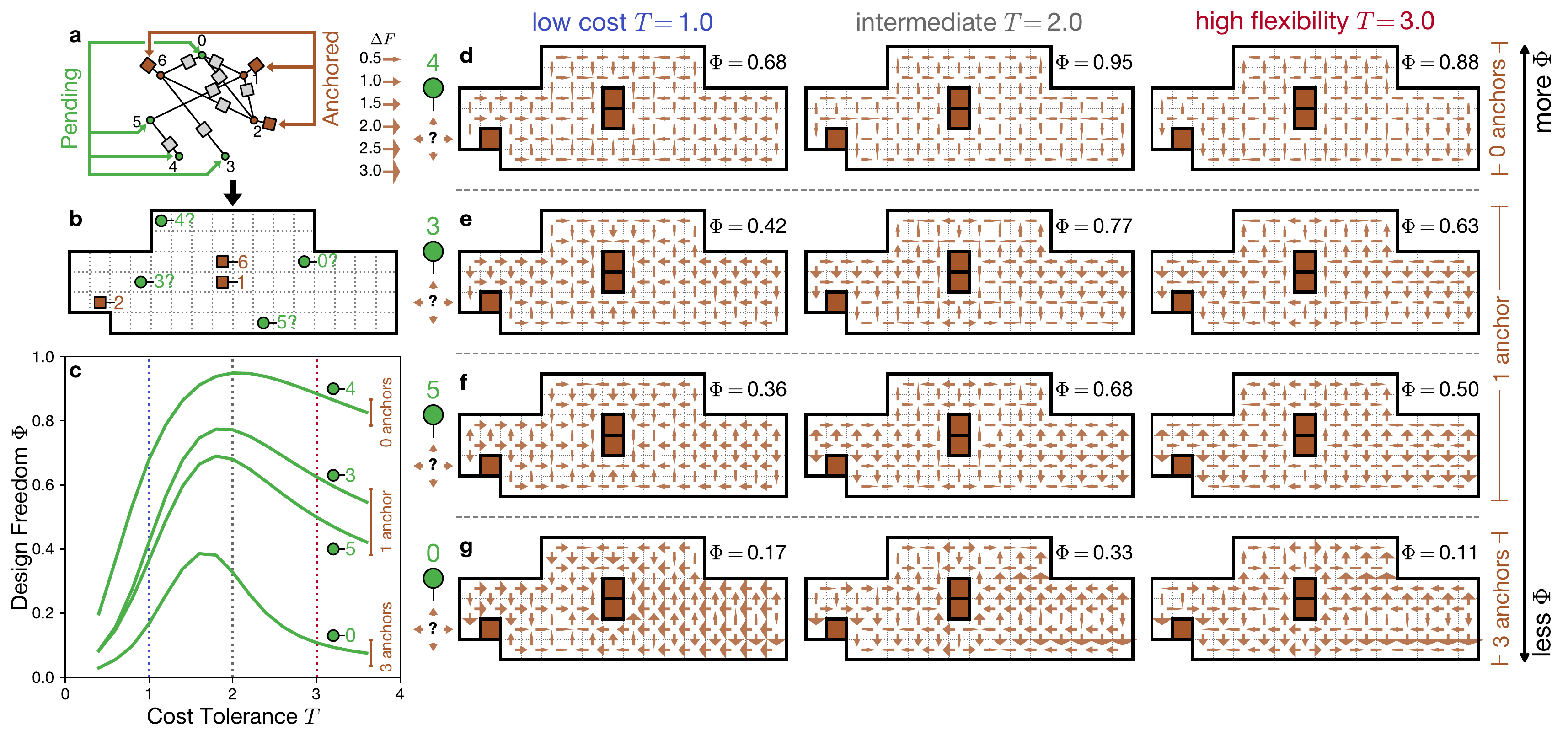}
  \caption{Early stage design decisions determine the association patterns and
  design freedom for subsequent ones. (a) Tensor network used for association
computations. Three out of seven units have already been anchored to specific
locations (brown), other four are pending placement (green). External legs not
shown. (b) The units 1,2,6 are anchored at the indicated locations within the
ship hull (brown squares). The units 0,3,4,5 can still be placed in many
locations, some demonstrated for an example (green circles). (c) Graph of the
design freedom $\Phi$ for the four units without anchors across a range of cost
tolerance $T$. Brown brackets on the right indicate that units with more
adjacent anchors have less remaining design freedom. Vertical dotted lines
indicate the $T$ values investigated in more detail in panels (d-g), as well as
in avoidance and adjacency patterns. (d-g) Design stress $\Delta F$ patterns for
the placement of each pending unit (rows in order of decreasing design freedom
$\Phi$) at three values of $T$ (columns, color coded). Legend for design stress
magnitude $\Delta F$ is shown to the left of panel (d).}
	\label{fig:robust}
\end{figure*}

Analyzing association patterns extends our avoidance and adjacency
investigations to situations in which multiple, pre-existing constraints
restrict functional units, i.e.\ in brownfield settings. These settings model
either of two situations: (i) actual late-stage design in which multiple
functional units have been fixed during preceding design stages, or (ii) an
early-stage design investigation of hypothetical late-stage situations under
different decision scenarios.

\emph{Constraints and Localization}---In either case, the expectation that
multiple active constraints will drive complex forms of interaction suggests
that identifying patterns of association that result will require different
techniques than identifying patterns of avoidance and adjacency. In general we
expect that patterns of association arising from multiple constraints will localize those
patterns relative to fixed design elements. This suggests that metrics of
association patterns should signal a tendency toward (or away from) placement
proximity relative to fixed elements, either globally or locally. Here, for a
global signal we adapt measures of emergent localization to compute a scalar
design freedom for unit locations. For a local signal we compute the design
stress associated with specified, hypothetical unit placement.

To study the effect of added constraints, their interplay with the logical and
physical architecture and the resulting localization, we employ the same model
system as in the avoidance and adjacency investigations. However we introduce
constraints that fix units 1,2, and 6 to specific locations. We investigate the
emergent localization of the other 4 units with two metrics via the global
design freedom $\Phi$ and local design stress $\Delta F$. Results are shown in
Fig.~\ref{fig:robust} and broken down below.

%and patterns of \emph{Landau free energy} (LFE) $F(\vec{x}_i)$.

\emph{Global Signal of Association: Design Freedom} Mathematically, both global
and local metrics of localization use a tensor network computation of
the marginal probability distribution $p(\vec{x}_i)$. The conversion of the
distribution into design freedom is inspired by the metric of existence area,
commonly used in studies of the Anderson localization of wavefunctions in
disordered media and the localization of vibrational
eigenmodes.\cite{thouless1974disordered,filoche2009localization} We define
design freedom as:
\begin{equation}
\Phi=\frac{1}{Y_0}\frac{\left( \sum\limits_{\vec{x}}p(\vec{x}) 
\right)^2}{\sum\limits_{\vec{x}}p(\vec{x})^2}\; ,
\label{eqn:localization}
\end{equation}
where the normalization $Y_0$ is the total number of cells within Physical
Architecture; in this example $Y_0=78$. Given this normalization, $\Phi$ takes a
value between 0 and 1 and has the meaning the effective fraction of the total
area available for unit placement, if the distribution was uniform. For a unit
with uniform distribution $p(\vec{x}_i)=const$, $\Phi$ would be 1, whereas for
an anchored unit $\Phi$ would be $1/Y_0 \to 0$.

Because of the heterogeneous connectivity of the logical architecture, design
freedom $\Phi$ varies between units. The variation between units is in addition
to variation with design pressure, via changing cost tolerance $T$.
Fig.~\ref{fig:robust}c plots $\Phi(T)$ by unit, and shows that all units have
design freedom peaks near $T\approx 2.0$. In the range of this near-critical
$T$, cost (effective attraction) and flexibility (effective repulsion) drivers
of unit interactions balance and allow the units to explore the largest range of
placement. As well, we observe $\Phi(T)$ to fall into three groups according 
to how constrained each unit is. Unit 4 is not directly connected to any of the
anchored units and thus enjoys the largest design freedom, almost approaching
$\Phi=1$. Units 3 and 5 are each connected to one anchored unit and thus have
intermediate $\Phi$. Unit 0 is connected to all three anchored units and thus
has the lowest $\Phi$ which quickly decays at both low and high $T$.

\emph{Local Signal of Association: Design Stress}---Whereas $\Phi$ serves as an
global scalar metric of design freedom, it is also important to understand how
global design freedom is distributed locally. This local distribution is
captured by design stress. Design stress is closely related to an effective
(Landau) free energy (LFE), defined as follows:
\begin{equation}
F(\vec{x}_i)=-\ln p(\vec{x}_i)+C\; ,
\end{equation}
where $C$ is an arbitrary additive constant. We chose a convention where $C$ is
such that the minimal value of $F$ is zero. The LFE can be interpreted as an
effective design objective for the chosen degree of freedom, given that other
degrees of freedom have been fixed or integrated out. This interpretation is
analogous to the void premium (Eq.\ \ref{eqn:void_cost}) in our avoidance
investigation, but instead of the effects of unit placement on voids, here we
examine the effects of unit placements on one another.  Similar to the
definition of void design stress via a spatial difference, the difference of LFE
between two horizontally or vertically adjacent cells is the design stress
$\Delta F$. Design stress is then an effective ``force'' that pushes individual
functional units towards their preferred locations.\cite{systemphys,
robustdesign} Compared with global design freedom, design stress patterns
give a more detailed picture of effective localization.

Fig.~\ref{fig:robust}d-g presents the design stress patterns for all four
pending units at three values of $T$. Design stress is represented by brown
arrows drawn across the boundary of two adjacent cells and pointing from higher
to lower LFE. In other words, to decrease LFE and reach lower values of its
effective design objective, a unit needs to follow the arrows towards a basin.
As the basin gets smaller and its walls get steeper, the pending units become
more closely associated with the anchored ones and thus exhibit stronger
emergent localization. The localization effect is strongest at lowest $T$, where
all of pending units are strongly attracted to the two anchored units 1,6 in a
single basin. The basin is steepest for the most constrained unit 0, less steep
for the units 3,5, and the shallowest for the least constrained unit 4,
consistent with our expectation based on design freedom $\Phi$. At higher
$T=2.0,\,3.0$, the constrained unit 0 develops complex LFE and design stress
landscapes with multiple local minima, maxima, and ridges (panel g). Units 3,5,
connected respectively to the anchored units 6 and 1 in the center of the hull,
show an X-shaped pattern of LFE (panels e,f), similar to the bond diagrams for
directly connected unit pairs, e.g. Fig.~\ref{fig:bonds1}d. Lastly, the unit 4
is not connected to any of the anchored units, instead it is ``dangling off''
unit 5 and thus shows almost nonexistent design stress across the whole hull
(panel d).

\emph{Association: Interaction and Decision Drivers}---Both the design stress
and design freedom metrics show that the association patterns and the emergent
localization phenomenon strongly depend on the position of both the fixed and
the pending units within the logical architecture. The logical architecture
alone gives an interpretation of the emergent localization result by counting
the anchored neighbors. However, fully predicting localization requires
examining the logical--physical architecture interplay that arises from the
systems physics analysis.  Unlike the simplified unidirectional design stress
discussed in Ref.~\cite{robustdesign}, in this system the design stress pattern
is emergent both from unit interactions and from previous design decisions.
Chaining design decisions into sequences and achieving optimal control of
emergent localization stands out as an important question for further study.

\section{Discussion}
\begin{figure}
\includegraphics[width=0.45\textwidth]{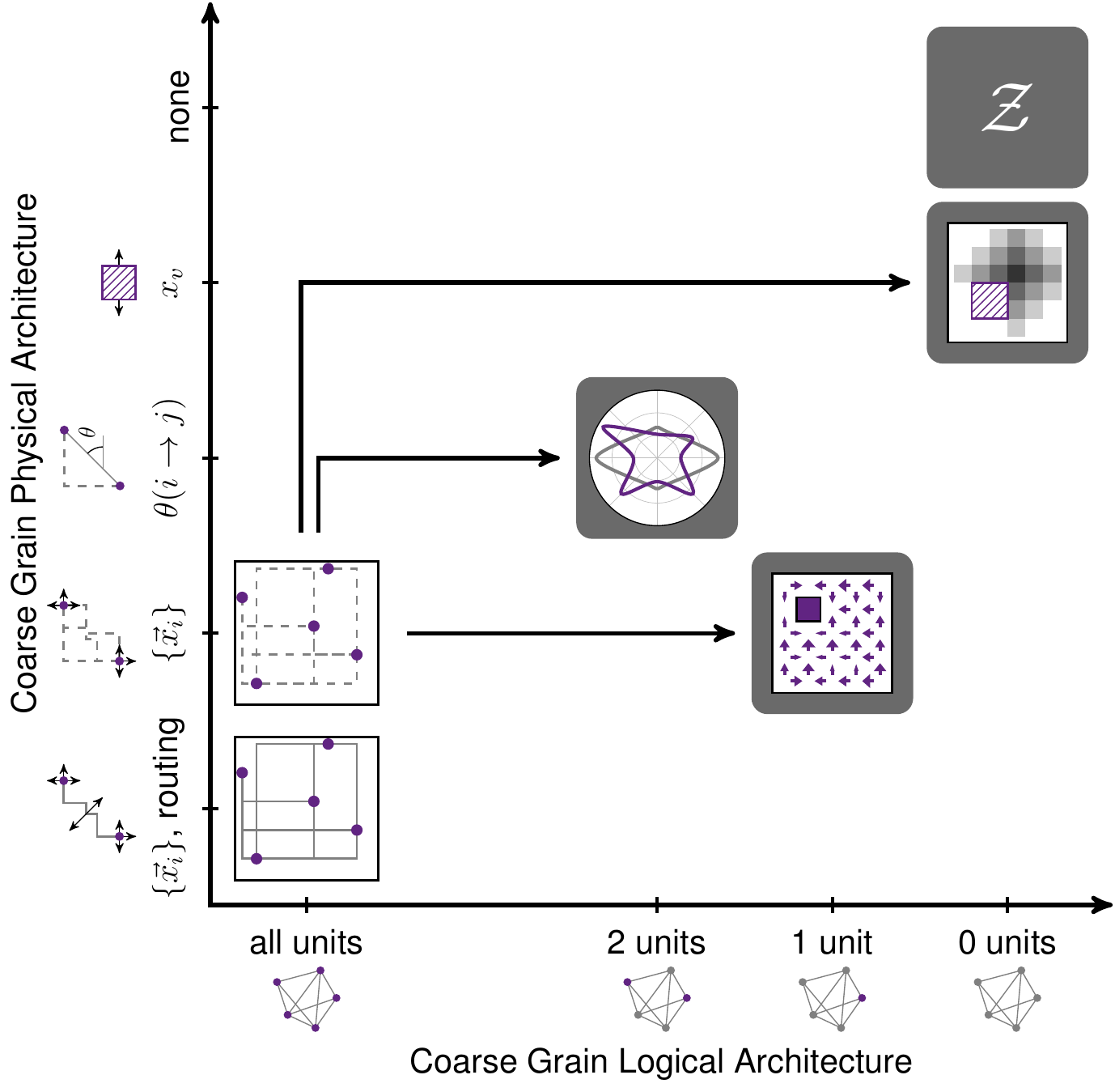}
\caption{System exploration across levels of detail in physical and logical 	architectures via two orthogonal directions of coarse-graining. Horizontal axis represents the reduction in the number of functional units explicitly considered. Vertical axis represents the reduction in spatial detail considered. Black arrows represent the computational pathway across the  study of avoidance, adjacency, and association patterns. The full design objective $\mathcal{O}$ (Eqn.~\ref{eqn:obj_fn}) depends on all unit locations and routings. The effective design objective $\mathcal{O}_\textsf{eff}$ (Eqn.~\ref{eqn:obj_fn_spec}) depends only on locations of all units. The association pattern considers detailed location $\vec{x}_i$, but only for a single unit. The adjacency pattern reduces the spatial detail to just the relative direction $\theta(i\to j)$ between two units. The avoidance pattern further reduces the spatial detail to just the void position $x_v$ as a single collective coordinate of the unit cloud. Finally, the partition function $\mathcal{Z}$ loses all detail of physical and logical architecture and summarizes the properties of the whole design space.
}
\label{fig:coarse}
\end{figure}

In this paper we showed that questions of avoidance, adjacency, and association
among the elements of complex, distributed systems hinge on the interaction
between logical- and physical-architecture description planes. We bridged these
descriptive planes with statistical physics techniques and showed that patterns
of avoidance, adjacency, and association can be mapped for an example system.

\emph{Design Phenomena: Symmetry-Breaking, Emergent Adjacency, Localization}---
Our mapping of avoidance gave a space premium landscape. We found this landscape
to undergoe a symmetry-breaking transition with a change from design pressure that
prioritizes high flexibility to pressure that prioritizes low cost
(Figs.~\ref{fig:void1},\ref{fig:void2}).
Our mapping of adjacency gave a description analogous to ``bond'' directions in
matter systems. From this bonding description we observed that indirectly connected
design elements developed emergent adjacency (Fig.~\ref{fig:bonds1}). We also
found large downstream changes in adjacency from small changes in underlying
connectivity (Fig.~\ref{fig:bonds2}).
Our mapping of association patterns quantified changes in global design freedom
driven by fixing design elements and changes in design pressure. Mapping these
effects locally showed the emergent localization of design elements
(Fig.~\ref{fig:robust}).

\emph{Coarse Graining for Other Design Contexts}---Our mappings of avoidance,
adjacency, and association patterns were done for a model system motivated by
problems in Naval Engineering. However, for other design contexts where
questions of avoidance, adjacency, and association patterns arise, our
statistical physics approach opens new lines of attack. In particular, our
approach can be summarized in two steps. First we ``decorated'' the logical
architecture with detail from the physical architecture. Then, we systematically
chose two sets of system details, one to examine in detail, and the other to
treat in aggregate, in a coarse-grained way. The aggregated details induce
effective patterns of interaction among the remaining elements, that reveal
underlying patterns of arrangement. We illustrate this strategy in
Fig.~\ref{fig:coarse}. Fig.~\ref{fig:coarse} casts the strategy into two
orthogonal forms of coarse-graining: one in the physical architecture, the other
in the logical architecture. In this representation, in statistical physics
language, microstates that retain complete detail of both the physical and
logical architecture sit in one corner, whereas the partition function, which
aggregates microstates into a single scalar sits in the opposite corner.
Though the specific locations that correspond to our investigations are given at
specific points on these axes, regardless of design context, answers to
questions about patterns of avoidance, adjacency, and association lie
at intermediate levels of detail between those extremes.

\section*{Acknowledgements}
We thank C.X.~Du for useful discussions, A.S.~Jermyn for extensive help with the 
\texttt{PyTNR} package, and N.~Mackay for assistance in development of the \texttt{TenZ} 
package. This work was supported by the U.S. Office of Naval Research Grant Nos.\
N00014-17-1-2491 and N00014-15-1-2752. GvA acknowledges the support
of the Natural Sciences and Engineering Research Council of Canada (NSERC).

\appendix

\section{Methods}
\label{sec:methods}
\subsection{Tensor Network Construction}
\begin{figure*}
\includegraphics[width=\textwidth]{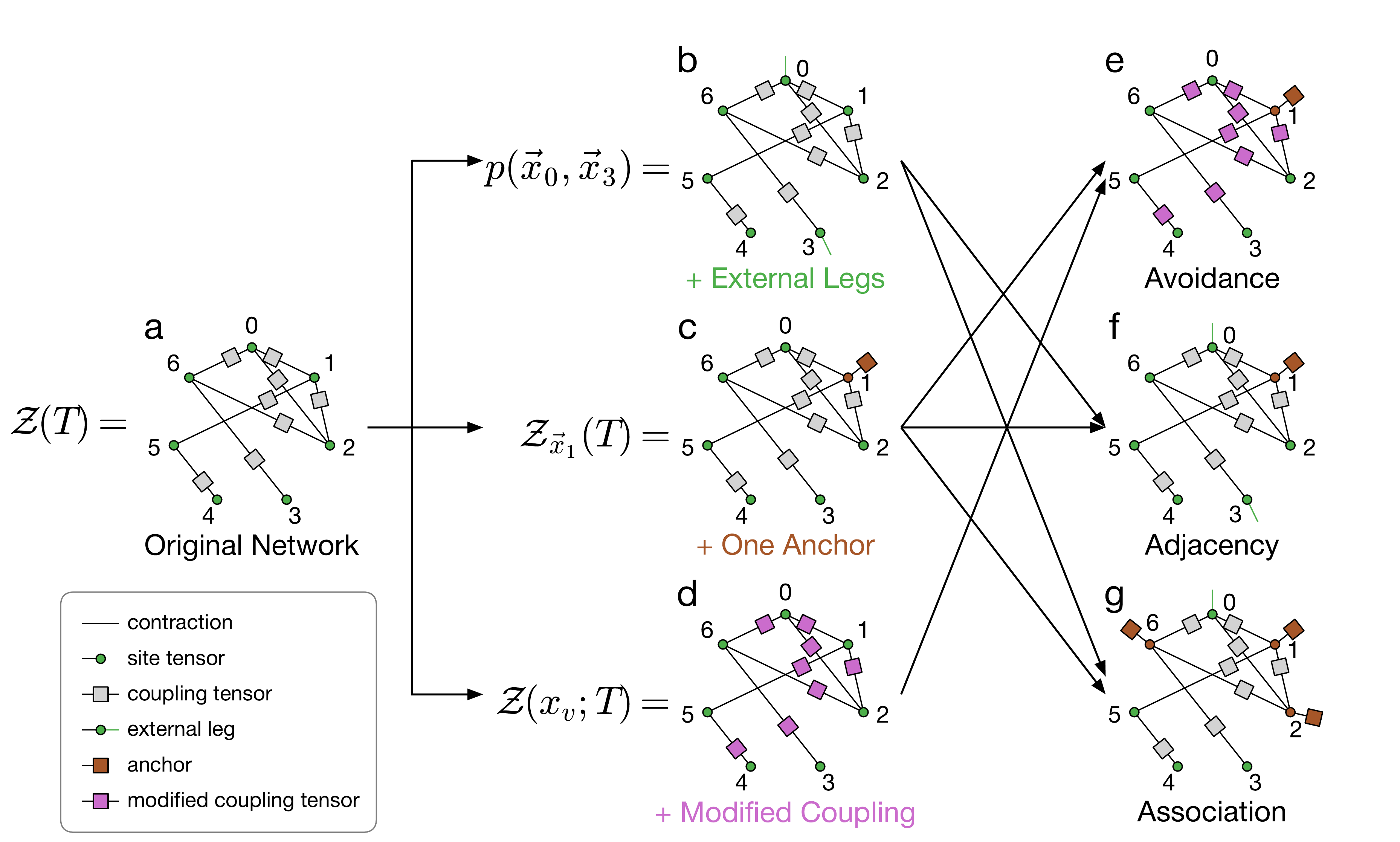}
\caption{Tensor networks can be used as information structures as combinations of basic moves express complex questions about the design space. Panel (a) shows the original network, panels (b-d) describe the three basic moves that modify its topology, panels (e-g) show how the elementary moves are recombined to study the emergent patterns. (bottom-left corner) Legend of tensor network elements. (a) The original network connects $n=7$ site tensors (green circles) with coupling tensors (gray squares) in the same pattern as the Logical Architecture (Fig.~\ref{fig:setup} top). Since this network has no outgoing legs, it contracts into a single scalar number equal to the partition function $\mathcal{Z}(T)$. (b) Move 1 adds extra outgoing legs on site tensors 0 and 3 (green lines), making the contraction result in the rank-2 tensor containing the joint marginal probability distribution on the spatial positions of the two units $p(\vec{x}_0,\vec{x}_3)$. (d) Move 2 attaches an additional rank-1 anchor tensor (brown square) to site 1, fixing it to a specific spatial location. This network contracts to the conditional partition function $\mathcal{Z}_{\vec{x}_1}$. (d) Move 3 modifies all of the coupling tensors (shown as pink squares), for example to account for the voids. This network contracts to the modified partition function $\mathcal{Z}(x_v;T)$. (e) For the avoidance pattern, we use both anchors and modified couplings to compute the placement opportunity cost via Eqn.~\ref{eqn:void_cost}. (f) For the adjacency pattern, we fix node 1 with an anchor and compute the 2-unit marginal distributions $p(\vec{x}_i,\vec{x}_j)$ for all possible pairs of external legs $i,j$, and further convert them into bond order diagrams. (g) For the association pattern, we encode the past design decisions with anchors on units 1, 2, 6 and study the 1-unit marginal distributions on each of the other units.
}
\label{fig:tensors}
\end{figure*}

The partition function of the system can be expressed in terms of the effective
design objective in the following factorized form:
\begin{equation}
\mathcal{Z}(T)=\sum\limits_{\{\vec{x}\}}
\prod\limits_{\substack{i<j:\\A_{ij}\neq 0}}e^{-f(\vec{x}_i,\vec{x}_j;T)}.
\label{eqn:factored_Z}
\end{equation}
We interpret the factorized partition function \eqref{eqn:factored_Z} graphically in 
the form of a tensor network (Fig.~\ref{fig:tensors}a). Like other networks, a
tensor network consists of nodes and links. Each node is a tensor with rank
equal to node degree, and the network links represent the pattern of tensor
contractions. We construct tensor networks following a recent prescription of
Ref.~\cite{Jermyn1}, in which the tensor networks are bipartite: they consist of
\emph{coupling tensors} and \emph{site tensors}, with each type only connected
to the other type as shown in Fig.~\ref{fig:tensors}a. Logical architecture
$A_{ij}$ is contained in the pattern of the coupling tensors, whereas physical architecture domain $\{\vec{x}\}$ serves as the index of all the
tensors, and the design objective determines the contents of the tensors.

The basic tensor network (Fig.~\ref{fig:tensors}a) represents the partition
function $\mathcal{Z}(T)$ and is merely an alternative, graphical way to express
Eq.~\ref{eqn:factored_Z}. Each multiplicative term in the partition function
becomes a rank-2 coupling tensor with elements defined as $M_{\vec{x}_i
\vec{x}_j}\equiv e^{f(\vec{x}_i \vec{x}_j;T)}$ (note elementwise rather than
matrix exponentiation). The site tensors have the form of a multi-dimensional
Kronecker delta with the rank corresponding to site's degree in the logical
network $A_{ij}$. For example, the unit 0 in Fig.~\ref{fig:tensors}a has three
network neighbors, and thus corresponds to a rank-3 tensor $\delta_{\vec{x}_{01}
\vec{x}_{02} \vec{x}_{06}}$. The indices mean the value of unit 0 coordinate
$\vec{x}_0$ that is ``presented'' to each network neighbor, in this 
case units 1, 2, and 6. The Kronecker delta ensures that each neighbor perceives
the unit 0 at the same location, while index summation ensures that all possible
locations are considered. The rank of site tensors can be adjusted for other
computations, as shown below. For brevity of notation, we suppress indices on
edges because the contraction pattern is dictated by the graphical notation.

Contracting all the tensors along the network links corresponds to performing
the sum in Eq.~\ref{eqn:factored_Z}. Since that sum has no free indices,  the
network of Fig.~\ref{fig:tensors}a has no external (unpaired) legs. Contraction
of the network preserves the number of external legs, resulting in a rank-0
tensor, a scalar number.  Since each coupling tensor implicitly depends on cost
tolerance $T$, the result of contraction is the $T$-dependent partition function
$\mathcal{Z}(T)$.

In order to compute quantities other than the partition function, we add minor
modifications of the tensor network. These modifications are described in the
graphical language of moves. Here we define three moves: adding external legs,
adding anchors, and modifying the coupling tensors (Fig.~\ref{fig:tensors}b-d).
These moves are recombined to create networks that address the patterns of
avoidance, adjacency, and association (Fig.~\ref{fig:tensors}e-g).

\subsection{Move 1: External Legs} The first move adds extra legs to specific
site tensors to control whether specific design degrees of freedom are
marginalized or not. If none of the degrees of freedom are marginalized, then
carrying out the multiplication but not the summation in the sum
\eqref{eqn:factored_Z} would result in an un-normalized joint probability
distribution $p(\{\vec{x}_i\})$ over all the units, which is a rank-$N$ tensor
of prohibitive size.  However, following the usual probability theory calculus,
in a joint probability distribution each of the entering variables can be in
three states: joint, marginalized, or conditional. In the tensor network
representation of Fig.~\ref{fig:tensors}a, \emph{every} variable
$\vec{x}_0,\vec{x}_1,\dots$ is marginalized, resulting in the distribution
normalization, i.e. the partition function $\mathcal{Z}(T)$.

In this perspective, a special action needs to be taken to \emph{not}
marginalize some of the variables. We do this by adding external legs to the
corresponding site tensors (green lines in Fig.~\ref{fig:tensors}b). External
legs are the site degrees of freedom that are \emph{not} summed over,
functioning as free indices for the sum \eqref{eqn:factored_Z}. The result of
contracting the network in Fig.~\ref{fig:tensors}b is a rank-2 tensor that
represents the un-normalized joint probability distribution
$p(\vec{x}_0,\vec{x}_3)$. Since the original network contracted to yield the
full $\mathcal{Z}$, the normalized probability distribution can be expressed as
$\tilde{p}=p/\mathcal{Z}$.

\subsection{Move 2: Anchors}
The second move adds anchor tensors to the network. The anchors represent the
design decisions already taken and woven into the information structure, thus
encoding the brownfield aspects of design. In tensor network language, this is
equivalent to fixing some of the local degrees of freedom $\vec{x}_i$ and thus
summing over a restricted ensemble, \emph{conditional} on the fixed $\vec{x}_i$.
We do this by creating an additional tensor that we call an ``anchor''. An
anchor is a rank-1 tensor (vector) that is coupled to a site tensor and is
illustrated as purple square in Fig.~\ref{fig:tensors}c. The elements of an
anchor vector are given by the Kronecker delta $\delta(\vec{x}_i,\vec{x}_a)$,
where $\vec{x}_i$ is the index connected to the site and $\vec{x}_a$ is the
specific location to which the functional unit is pinned as result of a design
decision. Since we didn't create any external legs, the tensor network in
Fig.~\ref{fig:tensors}c also contracts to a scalar number of \emph{conditional}
partition function $\mathcal{Z}_{x_6}$ that functions as a similar statistical
summary of the system as the original partition function $\mathcal{Z}$. 

\subsection{Move 3: Modified Coupling}
The third move modifies the coupling tensors $M(\vec{x}_i,\vec{x}_j)$ and traces
the effect of this modification on the partition function. In our study, we use
the modification to account for the void where no units can be placed. The
modification consists of suppressing the statistical weight of the void cells in
the coupling tensor:
\begin{equation}
M^* (\vec{x}_i,\vec{x}_j)= 
M(\vec{x}_i,\vec{x}_j)\prod\limits_{\vec{x}_v}(1-\delta(\vec{x}_i,\vec{x}_v))(1-
\delta(\vec{x}_j,\vec{x}_v)),
\label{eqn:coupling_void}
\end{equation}
where $\vec{x}_v$ denotes the positions falling into the excluded void. In the
network on Fig.~\ref{fig:tensors}d we modified each coupling tensor in this way
(marked in pink).  The network results in the modified partition function
$Z(x_v;T)$, from which we compute the void free energy via
Eq.~\ref{eqn:void_cost}.

\subsection{Bond Diagrams}
To compute bond diagrams, the raw two-unit distributions
$p(\vec{x}_i,\vec{x}_j)$ need to be converted into the angular distributions
$p_{i\to j}(\theta)$ in post-processing. Since all functional units are placed
in a discrete, finite, and fixed set of cells $\vec{x}_i$, we pre-compute the
directions between any pair of cells $\theta(\vec{x}_i,\vec{x}_j)$, measured in
radians from $0$ to $2\pi$, ahead of time and store them.

Within the design ensemble, the locations of units $\vec{x}_i$ are random, drawn
from the joint distribution encoded in the tensor network. We compute a series
of marginal two-units distributions $p(\vec{x}_i,\vec{x}_j)$ for all pairs $i<j$
(see the tensor network in Fig.~\ref{fig:tensors}f). Since the possible unit
locations are discrete, the possible directions $\theta(\vec{x}_i,\vec{x}_j)$ form
an artificially irregular discrete set. This numerical artifact would result in
a jagged direction distribution $p_{i\to j}(\theta)$. We smooth the distribution
by using a version of non-parameteric Kernel Density Estimation (KDE)
\cite{rosenblatt1956remarks,parzen1962estimation} with 
periodic boundary conditions in which higher angular harmonics are suppressed:
\begin{equation}
p_{i\to 
j}(\theta)=\frac{1}{\mathcal{N}}\sum\limits_{k}\sum\limits_{\vec{x}_i,\vec{x}_j} 
p(\vec{x}_i,\vec{x}_j)e^{-\frac{1}{2}(hk)^2}\cos \left( 
k(\theta-\theta(\vec{x}_i,\vec{x}_j)) \right)
\label{eqn:bond_kde}
\end{equation}
Here $\mathcal{N}$ is a normalization factor, $h$ is the KDE smoothing factor
(bandwidth), $k\in \{0,1,2,\dots\}$ is the angular mode index. We find that
using smoothing factor of $h=0.1$ radians and angular modes up to
$k_\textsf{max}=30$ gives good results.

The resulting distributions $p_{i\to j}(\theta)$ need to be compared with the null 
distribution induced by the Physical Architecture (ship hull shape). We compute
the null distribution by evaluating the formula \eqref{eqn:bond_kde} for
$p(\vec{x}_i,\vec{x}_j)=const$. The identical null distribution is shown in
every panel of Fig.~\ref{fig:bonds1}.

\subsection{Numerical Aspects of Computation}
Implementing the tensor networks described above on a computer requires two
different kinds of computational work: constructing the networks from Logical
and Physical Architecture and possible modifications with the three moves, and 
contracting said networks numerically. We perform these two tasks in Python.

Using code we developed, we create tensor networks by specifying the network
topology, the spatial domain geometry, the design objective, and additional
moves. These specifications are done via high-level commands, allowing for the
rapid generation of  diverse networks.

Tensor network contraction is handled by a Python package. Existing tensor
network packages use different methods of executing a sequence of pairwise
tensor contractions. The contraction result does not depend on the contraction
sequence, but the computational time and memory requirements rise by orders of
magnitude for suboptimal sequences. Optimal sequences are known for certain
frequently used networks, wherease for others one can use exhaustive enumeration
algorithms to find the optimal sequence and then execute it repeatedly for the
same network topology.\cite{pfeifer2014faster} However, almost all networks that
we contract in the present work are subtly different, and therefore might
require different contraction sequences. We perform all contractions with the
\texttt{PyTNR} package, an open-source general purpose tensor network
contractor.\cite{Jermyn1,jermyn2} The features of \texttt{PyTNR} include using
heuristics to automatically generate the contraction sequences on the fly, and
performing SVD approximations of controlled precision to reduce the dimension of
stored tensors.

The features of \texttt{PyTNR} define the computational constraints on the size
of systems that our approach can handle. The size and structure of the Logical
Architecture directly change the number of units $n$ in the network and the
number of tensors $n_t$ (counting both site and coupling tensors). The size or
resolution of the Physical Architecture domain directly affect the tensor bond
dimension $D$. A rigorous, though pessimistic upper bound on the time complexity
of contraction stands at $\mathcal{O}(D^{2\sqrt{\Delta n_t}})$, where $\Delta$ is the
maximal tensor rank.\cite{kourtis2018fast} In contrast, \texttt{PyTNR} relies on
a heuristic and stochatic generation of contraction sequences, which complicates
even the empirical investigations of numerical scaling of complexity. For highly
structured networks, such as $d$-dimensional hypercubic lattices, the time
complexity scales a power law $\mathcal{O}(N^\gamma)$, where the exponent $\gamma$ is
a bit larger than the space dimension $d$ and depends on the nature of system's
boundary conditions (periodic or closed).\cite{Jermyn1}

To provide more concrete numbers, each tensor network contraction in this paper
takes less than 10 seconds on a laptop computer (Intel Core i5-3360M @ 2.8GHz
CPU, 8Gb RAM) for our example system ($n=7$ units, $n_t=15$ tensors, $D=78$,
total number of combinatorial states $\mathcal{O}(10^{13})$). The example system size
was chosen to best illustrate the physical phenomena at single unit resolution.
In other investigations we reliably contracted lattice networks of up to
$n_t=\mathcal{O}(10^3)$ tensors accounting for $\mathcal{O}(10^{167})$ combinatorial
states using \texttt{PyTNR}.\cite{elo} This result suggests that the current
tensor network methods would remain tractable for systems even one order of
magnitude larger, or perhaps even larger as the tensor network methods develop.

\section{Supplementary Results}
\label{sec:supres}
\subsection{Avoidance: Excess Density}
\begin{figure*}
	\includegraphics[width=\textwidth]{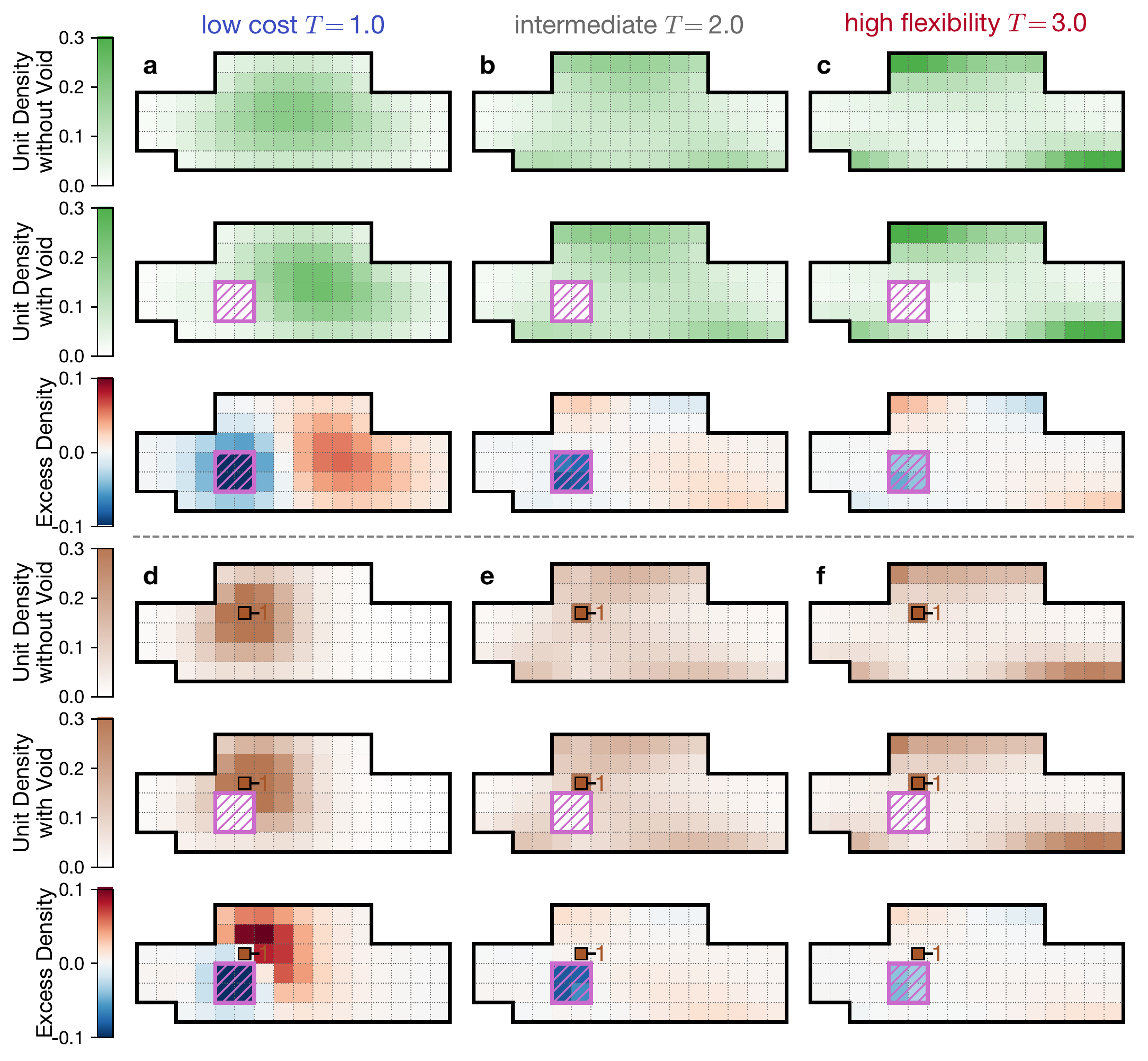}
	\caption{
		Reserving space in locations with large void premium causes large 
		rearrangements of the unit cloud morphology. The rearrangement is shown 
		via the density profile without the void, with the void, and their 
		pointwise difference (rows) for three values of $T$ (columns, color 
		coded). Panels (a-c) show the rearrangement in a fully greenfield  
		scenario (no anchors). Panels (d-f) show the rearrangement in a 
		brownfield scenario (one anchor indicated with a brown square).
	}
	\label{fig:void_excess}
\end{figure*}

In the main text of the paper we show that void premium quantifies the cost of 
reserving space within the ship hull, with the effect being further amplified 
by the presence of an anchor (Figs. 4-5). We associated high void premium with 
a large rearrangement of the functional units. We can quantify the degree of 
rearrangement by computing the unit density profile without void 
$\rho_\textsf{no void}(\vec{x})$, the unit density profile with void 
$\rho_\textsf{void}(\vec{x})$, and their difference: 
\begin{equation}
\Delta \rho (\vec{x})=\rho_\textsf{void}(\vec{x}) - \rho_\textsf{no 
void}(\vec{x}),
\end{equation}
which we term \emph{excess density} that can be both positive and negative. 
Since the total number of functional units does not change upon addition of 
void, the positive and negative regions of excess density have to cancel each 
other. In this case large rearrangements of the unit cloud are characterized by 
large \emph{contrast} between the positive and negative regions, graphically 
visible in saturation of colors.

We plot all three densities in Fig.~\ref{fig:void_excess}, both without and 
with an anchor on unit 1. In low-cost case $T=1.0$ (panel a) the units want to 
form a compact cloud, which can be located anywhere within the hull. Upon 
creation of a void slightly to the left of center, the unit cloud relocates to 
the right of the void, as seen by large negative $\Delta \rho$ in the left half 
of the hull and positive in the right half (visible as red and blue clouds). 
When unit 1 is anchored (panel d), this effect becomes even more pronounced 
since the unit cloud condenses around a reference point. Creation of a void 
pushes the unit cloud to the right and above the anchor, but it cannot move far 
from the anchor, resulting in high contrast of excess density (strong color 
saturation in the figure) and thus high void premium. At intermediate and high 
values of $T$, both with and without an anchor (panels b,c,e,f) the 
rearrangements are much smaller, visible in much paler colors on excess density 
heatmaps.

\subsection{Adjacency: All Bond Diagrams}
\begin{figure*}
	\includegraphics[width=\textwidth]{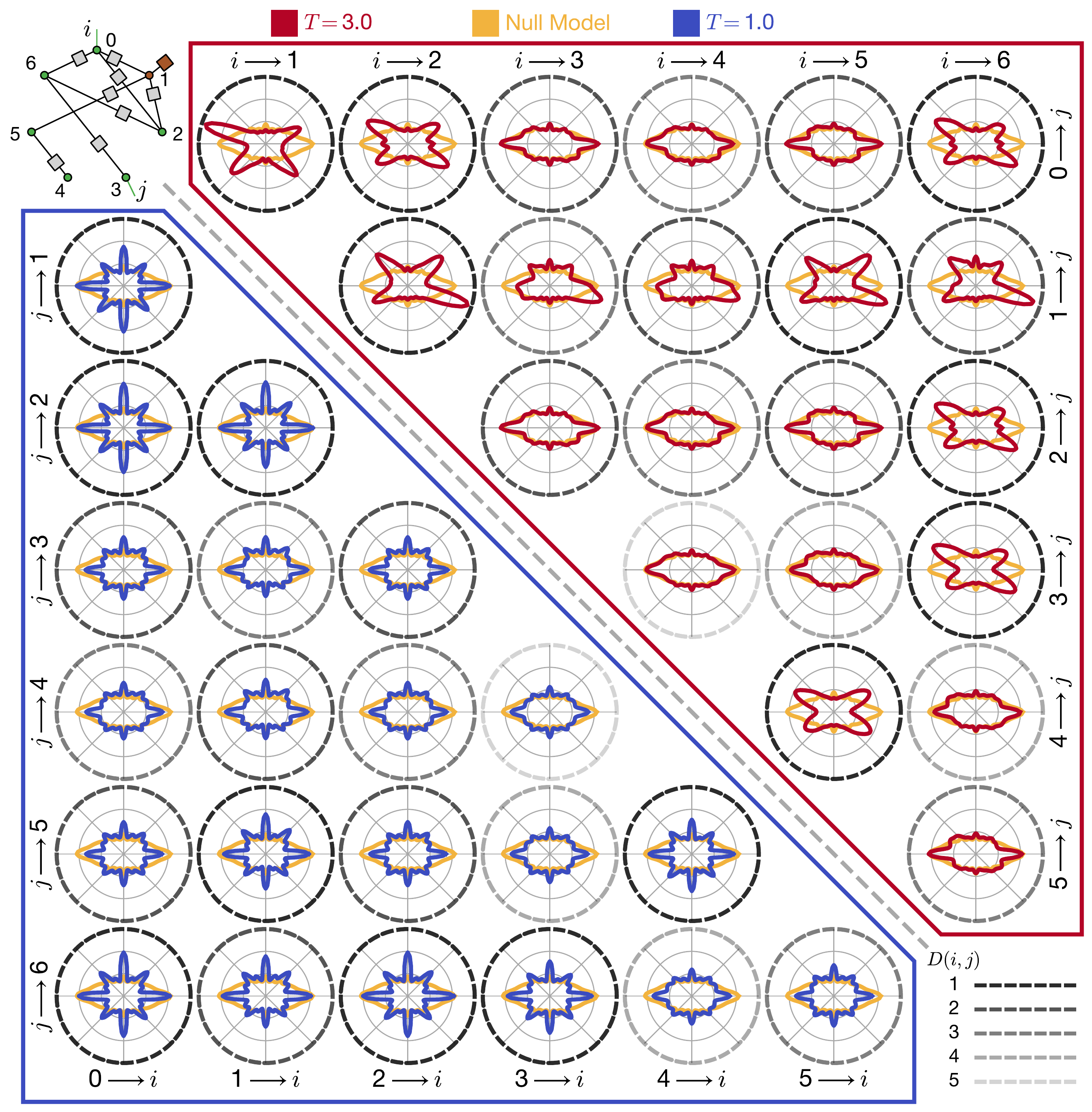}
	\caption{
		Adjacency patterns between the 7 functional units shown via bond 
		diagrams $p_{i\to 
			j}(\theta)$. (top-left corner) Typical tensor network used for bond 
			diagram 
		computations, with a brown anchor for unit 1 and green external legs at 
		units $i$ and 
		$j$ for each origin-destination pair $i,j$. (bottom-right corner) 
		Legend for the 
		opacity of axes boundaries representing the topological distance 
		$D(i,j)$ varying from 
		1 to 5 hops. (top-right triangle, red) Bond diagrams for $T=3.0$. 
		(bottom-left triangle, 
		blue) Bond diagrams for $T=1.0$. The origin and destination of each 
		bond diagram are 
		indicated on the outside boundaries of the triangle. Axes in positions 
		symmetric with 
		respects to the diagonal refer to the same origin-destination pair and 
		thus can be directly compared. The yellow curve in 
		each axes shows the null model bond diagram $p_0(\theta)$.
	}
	\label{fig:bonds_full}
\end{figure*}

In the main text of the paper we show how to compute the bond diagram $p_{i\to 
j}(\theta)$ for any pair of units $i,j$. Since the directions $\theta(i\to j)$ 
and $\theta(j\to i)$ only differ by a trivial rotation by angle $\pi$, and the 
direction from a node to itself is not defined, a system of $n$ units would 
have $n(n-1)/2$ independent bond diagrams. The Logical Architecture of the 
example problem was deliberately chosen to not have any graph symmetries, 
therefore the bond diagrams are not related to each other via any symmetries.

While in the main text we only show several representative bond diagrams 
(Figs.~6-7), Fig.~\ref{fig:bonds_full} shows all of the bond diagrams for the 
low-cost regime $T=1.0$ and the high-flexibility regime $T=3.0$. The diagrams 
are arranged as a lower-triangular and an upper-triangular matrix of polar 
plots so that the diagrams in positions symmetric with respect to the diagonal 
refer to the same pair of functional units and are thus directly comparable. 
The full set of diagrams shows all adjacency features highlighted in the main 
text. For units that are directly connected, most of the diagrams at $T=1.0$ 
show eightfold signal (e.g.\ 2$\to$6), while diagrams at $T=3.0$ show X-shaped 
fourfold signal (e.g.\ 4$\to$5). All diagrams to or from unit 1 show further 
symmetry breaking because unit 1 is anchored in space. For units that are 
connected only indirectly with large topological distance, the bond diagram 
approaches the null model at both $T=1.0$ and $T=3.0$ (e.g.\ 3$\to$4).

%\bibliographystyle{apsrev4-1} 
%\bibliography{gmaster} 


\end{document}